\documentclass[review,12pt,authoryear,3p]{elsarticle}

\usepackage{amssymb}

\usepackage[utf8]{inputenc}
\usepackage{csquotes}
\usepackage{enumerate}
\usepackage[shortlabels]{enumitem}
\usepackage[T1]{fontenc}
\usepackage{url}
\usepackage{graphicx}
\usepackage{xcolor} 
\usepackage{amsmath}
\usepackage{amssymb}
\usepackage{csquotes}
\usepackage{multirow}
\usepackage{tabularx}
\usepackage[colorlinks]{hyperref}
\usepackage{tikz}
\usepackage{caption}
\usepackage{subcaption}
\usepackage{enumitem} 
\usepackage{textcomp}

\newcommand{\Mod}[1]{\ \mathrm{mod}\ #1}
\DeclareMathOperator*{\argmin}{arg\,min}

\journal{arXiv}

\begin{document}

\begin{frontmatter}

\title{Designing organizations for bottom-up task allocation: The role of incentives}

\author{Stephan Leitner}
\ead{stephan.leitner@aau.at}

\address{University of Klagenfurt, Universit\"atsstra{\ss}e 65-67, 9020 Klagenfurt, Austria}

\begin{abstract}
In recent years, various decentralized organizational forms have emerged, posing a challenge for organizational design. Some design elements, such as task allocation, become emergent properties that cannot be fully controlled from the top down. The central question that arises in this context is: How can bottom-up task allocation be guided towards an effective organizational structure? To address this question, this paper presents a novel agent-based model of an organization that features bottom-up task allocation that can be motivated by either long-term or short-term orientation on the agents' side. The model also includes an incentive mechanism to guide the bottom-up task allocation process and create incentives that range from
altruistic to individualistic. Our analysis shows that when bottom-up task allocation is driven by short-term orientation and aligned with the incentive mechanisms, it leads to improved organizational performance that surpasses that of traditionally designed organizations. Additionally, we find that the presence of altruistic incentive mechanisms within the organization reduces the importance of mirroring in task allocation.
\end{abstract}


\end{frontmatter}


\section{Introduction}
\label{sec:intro}

In recent years, there has been a trend towards less hierarchical organizational forms in both research and corporate practice \citep{vergne2020,lee2017,ben2016,puranam2014,zarraga2003}. This trend has given rise to a number of corresponding concepts. One such concept is that of holacracies \citep{robertson2015}, which describes organizations composed of multiple teams with decentralized management structures. In holacracies, individuals can be members of multiple teams, hold multiple roles, and exercise leadership functions. This flexible design promotes individualization and responsiveness within the organization \citep{bernstein2016,kumar2018}.
Another concept is that of self-managing organizations, in which employees are granted more rights than they are in holacracies \citep{lee2017}. In self-managing organizations, decision-making authority is radically decentralized, formally and systematically. As a result, every employee has a certain degree of decision-making authority, and once decisions are made, they cannot be overruled by others. Self-managing organizations are related to the concepts of boss-less organizations \citep{puranam2015,burton2017} and self-organizing organizations \citep{daft1993}, which allow employees to make both strategic and tactical decisions at all levels of the organization, including initiating new projects and self-selecting to work on them \citep{puranam2018,raveendran2022}. These recently emerging concepts belong to the class of evolutionary approaches to organizational design. These approaches differ from traditional ones in that they view organizational design as an emergent property rather than the result of intentional planning \citep{tsoukas1993,lawrence1967,galbraith1974}. 

The study of organizational design has a long history, with early research focusing on bureaucracies and multi-divisional organizations \citep{chandler1969,weber2009}. Later, contingency theory introduced the concept of differentiation based on contextual factors and emphasized the importance of internal coherence in organizational structures \citep{miller1984}. The related literature discusses a range of organizational design elements, including objectives, technology, participants, social structure, and the environment \citep{scott1998}. \cite{baligh1996}, for example, distinguish between contingency factors (e.g., technology, strategy, environment) and organizational design parameters, with the latter including organizational structures and their properties such as complexity, rules, and procedures. \cite{daft2015} argue that size and culture are also relevant design parameters and particularly focuses on the structural properties of organizations. In a similar vein, \cite{gebauer2010}, among others, focus on culture, structure, and human resource management practices.  \cite{good2019} review the literature on organizational design elements and argue that there is a substantial overlap in previous attempts to identify these elements. They conclude that there are a number of elements that are consistently considered important. In particular, the \textit{organizational purpose} (e.g., codified in objectives) is their starting point, from which the following three additional design parameters are derived: 
\textit{(i)} The \textit{activities} the organization must perform to fulfill its purpose, 
\textit{(ii)} the \textit{structure} that supports the performance of these activities, and 
\textit{(iii)} the \textit{people} who belong to the organization, the \textit{incentive mechanisms} to induce certain behaviors, and the organization's internal \textit{culture}. 
This paper focuses on \textit{(ii)} structures, with a particular emphasis on the division of labor, i.e., how tasks are divided between individuals inside the organization \citep{good2019}, and \textit{(iii)} incentive mechanisms to guide individual behavior. The \textit{(i)} activities required to fulfill the organization's purpose are considered fixed in this paper, meaning that organizations cannot affect them. However, the paper takes into account activities of different complexity.

Previous research has shown that a good fit between organizational design elements is crucial for performance, while a misfit often leads to disorganization and lower performance \citep{schlevogt2002,donaldson2014,baligh1996}. Much of the previous research on organizational design has focused on this fit from a top-down perspective, in which organizational design elements are planned by a central designer before implementation. However, this paper focuses on \textit{designing organizations for bottom-up task allocation}, in which the structure becomes an emergent property of the organization rather than an element that can be planned top-down. As such, a central question arises: How can the process of bottom-up task allocation be guided towards an effective structure, in the spirit of guided self-organization \citep{prokopenko2009}?
This study aims to contribute to this body of research and advance the understanding of emergent organizational design by exploring the relationship between bottom-up task allocation, incentive mechanism, and organizational performance. In particular, the following research questions will be addressed:
\begin{enumerate}
\item[(1)] How do bottom-up task allocation mechanisms and incentive mechanisms interact and influence organizational performance?
\item[(2)] To what extent is the structure that emerges from bottom-up task allocation modular and how does this affect organizational performance?
\item[(3)] How does task complexity impact the results observed in relation to the interactions between bottom-up task allocation and incentive mechanisms and modularity?
\end{enumerate}
To answer these questions, we present a novel agent-based model of an organization that features bottom-up task allocation and includes various incentive mechanisms to guide the behavior of agents. Through our analysis, we find that bottom-up task allocation motivated by short-term objectives and aligned with the appropriate incentive mechanisms leads to improved organizational performance, even outperforming traditionally designed organizations. Additionally, we discover that the presence of certain incentive mechanisms within the organization can reduce the importance of mirroring in task allocation. By using this model, we aim to address the challenges of guiding bottom-up task allocation towards an effective organizational structure.

The structure of this paper is as follows: In Sec. \ref{sec:background}, the concept of mirroring in task allocation is discussed. In Sec. \ref{sec:model}, we introduce the agent-based model, detailing the simulation setup and data analysis. The results of the simulation are presented and discussed in Sec. \ref{sec:results}. Finally, we summarize and conclude the paper in Sec. \ref{sec:conclusion}.

\section{Mirroring in task allocation}
\label{sec:background}

Organizational design research often recommends following the \enquote*{mirroring hypothesis}, which suggests that an organization's formal \textit{structure} (in terms of task allocation) should reflect the technical characteristics of the tasks it faces, particularly in terms of interdependencies \citep{sanchez1996,langlois1992,colfer2016}. To align with this hypothesis, organizations may consider implementing a modular structure that minimizes dependencies between modules \citep{lawrence1967}. 
The mirroring hypothesis can be grounded in the concept of complex systems \citep{simon1991}. These systems consist of many components that often interact nonlinearly \citep{langlois2002}. When designing such systems, modularity---decomposing the system into interdependent modules with decoupled interfaces \citep{ulrich1995}---is a useful strategy for managing complexity. This logic can be applied to organizational design, where modularity is a spectrum, with full modularity (independent modules) at one end and integrality (no decomposition or recombining of modules into one system) at the other \citep{schilling2000,chen2017}.

For the context of product design, \cite{ulrich1995} provides a more detailed definition of modular design, including modularity at the level of functional components and interfaces between modules \citep[see also][]{schilling2000,sanchez1996,sanchez2013}. \cite{peng2018} also distinguishes between two levels of modularity: component modularity (the independence of a module within a complex system) and product modularity (the modularity of the entire system). They argue that even if some modules are highly decoupled, the system as a whole may not be modular if there are any interdependencies between other modules. Modular product architecture creates an information structure that is self-contained within the modules, and from an organizational design perspective, this structure allows for creating organizational units that reflect the product architecture and enable coordination within decoupled units. This can decrease the need and cost of coordination \citep{colfer2016}, making modularity a promising long-term organizational design strategy as long as the underlying product or task characteristics do not change.

The concept of the mirroring hypothesis is also rooted in transaction cost economics (TCE) \citep{williamson1985}. \cite{chen2019} argue that conflict resolution and information exchange are easier and more cost-effective within a firm than between firms, so the costs of governing transactions are higher when interdependent tasks are allocated to different firms than when they are allocated to the same firm.\footnote{Note that, depending on the scope of the analysis, this argument can be applied to units within a firm as well as to firms.} However, the administrative costs of hierarchical governance within a firm are likely to be lower than the corresponding transaction costs. Therefore, organizations should allocate interdependent tasks to agents \textit{within} a single firm to optimize transaction and administrative costs. Thus, TCE suggests that the boundaries of organizations should be set to minimize transaction costs  \citep{baldwin2008}, and the mirroring hypothesis states that these boundaries should reflect the technical characteristics of the underlying tasks \citep{colfer2016}.

Previous research has provided support for the mirroring hypothesis in various contexts.
For example, \cite{cabigiosu2012} found a positive correlation between the coupling of product and organizational architectures in the air-conditioning industry. \cite{tee2019} showed that modularization can help overcome coordination problems but may hinder collaboration in project settings. \cite{wei2021} analyzed mirroring in Chinese multinational enterprises and identified the boundary conditions and performance effects of different organizational archetypes. \cite{alochet2022} and \cite{chen2019} found evidence for partial mirroring (\enquote*{misting}) in organizations producing electric vehicles. Previous research has also supported misting as an effective strategy in industries with changing product architectures \citep{kosaka2021,burton2020}. 

However, not all studies support the mirroring hypothesis. \cite{colfer2016} reviewed 142 empirical studied and found that $70\%$ of the descriptive studies (focusing on the ties between technical task characteristics and organizational structures supported the mirroring hypothesis, $22\%$ provided partial support and $8\%$ rejected it. Their analysis of normative studies (focusing on the performance of mirrored and unmirrored architectures) showed that partial mirroring is superior in industries with dynamic technologies. However, $56\%$ of the analyzed studies on collaborative projects did not support the mirroring hypothesis. 
\cite{colfer2016} suggested that new forms of coordination due to technological advances may explain these results. Similarly, \cite{sanchez2013a} argue that \textit{(i)} cognitive, risk and capability, and commitment and discipline factors may explain why organizations fail to adopt the mirroring hypothesis.

\section{The agent-based model}
\label{sec:model}

\subsection{Model overview}

The main aim of this paper is to study organizations with bottom-up allocation of decision-making tasks (e.g., operational or procurement decisions) and and to analyze the effects of the incentive mechanism in place in the organization as well as task complexity on performance. The concept of decision-making tasks is abstract and can be applied to other domains, such as social interactions and team resilience \citep{massari2021} or product development \citep{ma2005}.\footnote{In this paper, the terms \enquote*{task} and \enquote*{decision} are used interchangeably to refer to decision-making tasks.}

The model considers \textit{agents} who represent organizational departments comprised of human decision makers, and together, the agents represent an organization with decentralized decision making. This decentralized structure is in line with the recently emerging organizational concepts, such as self-organizing or self-managing organization. Coordination across decision-making agents is solely facilitated through the incentive mechanism. Inspired by literature on information processing in organizations, the model also assumes that physical constraints make communication too costly \citep{marschak1972}. However, the model assumes that agents still behave selfishly, which is why coordination is required to assure coordinated actions across departments, and coordination through incentives is a feasible to do so \citep{fischer2008}.

The agents are characterized by limitations such as limited time and cognitive capacity. These limitations prevent them from solving the complex decision problem on their own, so they form a group to tackle the problem together. The agents have full autonomy regarding the division of labor. This means, they can autonomously make decisions on how to allocate sub-tasks, and they can also revise the task allocation over time. They are aware that there might be interdependencies between decision-making tasks, but they do not have complete information on the exact interdependence structure. However, they have the ability to learn about these missing pieces of information over time. 
The model considers two types of agents: those who make short-sighted decisions, focusing on immediate utility maximization without considering long-term effects, and those who consider long-term effects when allocating tasks. Agents of the latter type aim to minimize the interdependencies between sub-tasks assigned to different agents and maximize interdependencies within their own areas of responsibility, which is in line with the mirroring hypothesis. It is expected that optimizing for interdependencies will be beneficial in the long run. 

\begin{figure}
\centering
\includegraphics[width=0.7\textwidth]{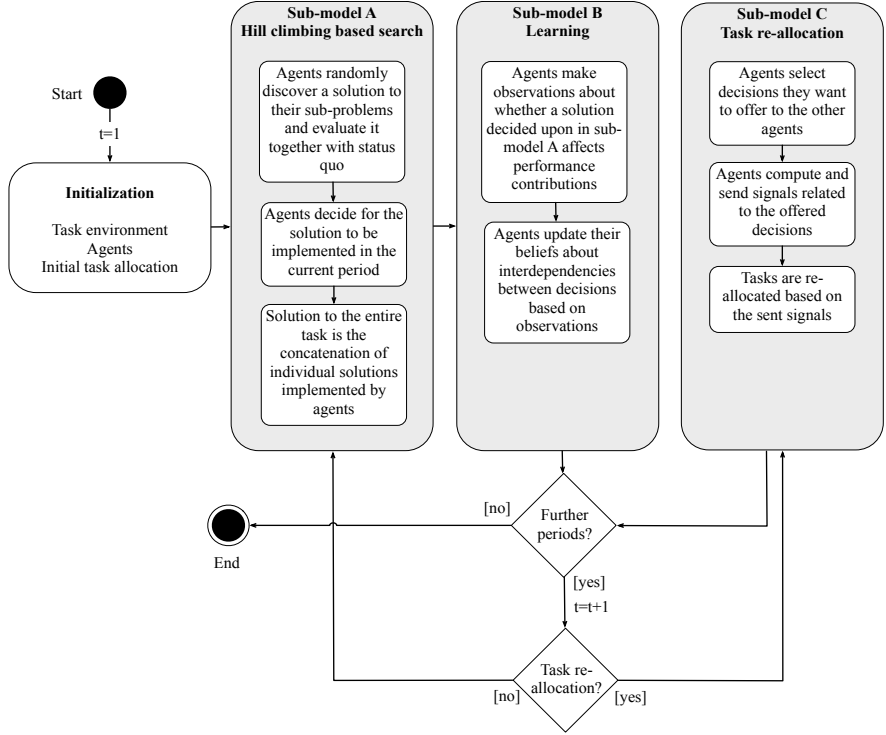}
     \caption{Model architecture and sequence of events}
     \label{fig:flowchart}
\end{figure}

Figure \ref{fig:flowchart} gives information on the model structure and sequence of events during the simulations. After initializing the performance landscape (Sec. \ref{sec:task-env}) and agents as well as the initial task decomposition (Sec. \ref{sec:decomposition}), agents begin a hill-climbing search for solutions to their partial decision problems (Sec. \ref{sec:hillclimbing}) and learn about the complexity of the task (Sec. \ref{sec:learning}). Every $\tau \in \mathbb{N}$ periods, agents are given the possibility to autonomously adapt the current task allocation (Sec. \ref{sec:reallocation}). We observe the overall task performance and the task allocation resulting from the autonomous allocation process for $t\in\{1,\dots,T\} \subset \mathbb{N}$ periods. The simulation model is implemented in Matlab\textsuperscript{\tiny\textregistered} R2022a. 

\subsection{Task environment}
\label{sec:task-env}

The model of a stylized organization builds on the $N\!K$-framework \citep{levinthal1997,wall2021}. An organization consists of $M\in \mathbb{N}$ agents who face a complex decision problem. We denote the decision problem by the $N$-dimensional vector
\begin{equation}
    \mathbf{d}=\left(d_1, \dots, d_N \right)~, 
\end{equation}
where $d_n \in {0,1}$ for $n \in {1, \dots, N}$.\footnote{For readability, the notion of $t$ is suppressed in Sec. \ref{sec:task-env}.} The number of solutions to the overall problem is $2^N$ and each solution is an $N$-digit bit-string. There are at most $K \leq N-1$ interdependencies between the decisions $d_n$, which means that the contribution of a decision $d_n$ to the task performance is affected by at most $K$ other decisions. This relationship can be formalized in the payoff function
\begin{equation}
\label{eq:perf-contribution}
c_n = f\left(d_n, d_{i_1}, \dots, d_{i_K}\right)~,
\end{equation}
where $\{i_1, \dots, i_K\} \subseteq \{1, \dots, n-1, n+1, \dots, N\}$. The performance contributions are independently drawn for the uniform distribution so that $c_n \sim U\left(0,1\right)$. The overall task performance for a solution $\mathbf{d}$ is the mean of the individual performance contributions $c_n$:
\begin{equation}
\label{eq:task-perf}
    c(\mathbf{d}) = \frac{1}{|\mathbf{d}|} \sum_{n=1}^{|\mathbf{d}|} c_n~,
\end{equation}
where the function $|\cdot|$ returns the length of a vector. 

The mapping of solutions to the decision problem $\mathbf{d}$ and their corresponding performances, as defined in Eqs. \ref{eq:perf-contribution} and \ref{eq:task-perf}, is used to create performance landscapes. The interdependencies between decisions shape the complexity of the decision problem in terms of the ruggedness of the resulting landscapes. As the number of interdependencies $K$ increases, the number of peaks (and local maxima) increases. For example, if $K=0$, the landscape is smooth the global maximum is relatively easy to find. In contrast, if $K=N-1$, the landscape is maximally rugged landscape with numerous local maxima, and it becomes more difficult to search for the global optimum.\footnote{In this case, the expected number of global optima is $2^N/(N+1)$. For landscapes with intermediate values of $K$, the number of peaks interpolates between the extreme cases of $1$ and $2^N/(N+1)$ \citep{kauffman1993,tomassini2008}.}  

\subsection{Agents and task decomposition}
\label{sec:decomposition}

Agents in the organization have limited capabilities and/or resources, such as  limited cognitive capacities, time, or other resources to solve the entire $N$-dimensional decision problem alone. Therefore, they need to collaborate and work together to find solutions to the problem. Let us denote the maximum number of decisions that an agent can handle at a time by $Q\in\mathbb{N}$. To prevent agents from dropping out of the group, every agent must be responsible for at least one decision at a time. This means that $1\leq Q < N$.

Agents decompose the decision problem $\mathbf{d}$ into $M$ disjoint sub-problems. We denote the decisions in agent $m$'s area of responsibility at time $t$ by 
\begin{equation}
\mathbf{d}_{mt} = [d_{j_i},\dots, d_{j_Q}]~,
\end{equation} 
where $\{j_1, \dots, j_Q\} \subset \{1, \dots, N\}$ and  $m\in\{1,\dots,M\}\subset\mathbb{N}$. The complement of $\mathbf{d}_{mt}$ in $\mathbf{d}$ is referred to as agent $m$'s residual decisions in period $t$: 
\begin{equation}
    \mathbf{d}_{-mt} = \mathbf{d} \setminus \mathbf{d}_{mt}
\end{equation}
Initially, the sub-tasks are allocated sequentially and symmetrically to agents, so that in period $t=0$, every agent is in charge of the same number of decisions, i.e., $|\mathbf{d}_{m0}| = M/N$. During the simulations, agents can adapt the task allocation following the procedure described in Sec. \ref{sec:reallocation}. We assume hidden action during the decision making procedure, which means that agents are always aware of the solutions to their sub-problem $\mathbf{d}_{mt}$. However, the decisions taken by other agents, i.e., the solutions to the residual decision problem $\mathbf{d}_{-mt}$, can only be observed in $t+1$, after they have been implemented.

The agents derive utility from the implemented solutions to the decision problem (Sec. \ref{sec:hillclimbing}). The organization uses a linear incentive scheme to reward agents for their contributions. For each agent $m$, the incentive scheme distinguishes between the contribution to task performance generated from the decision in the agent's area of responsibility and the residual performance. Agent $m$'s utility function is given by:
\begin{equation}
\label{eq:utility}
    U_{}(\mathbf{d}_{mt},\mathbf{d}_{-mt}) = a \cdot c\left( \mathbf{d}_{mt} \right) + (1-a) \cdot  c\left( \mathbf{d}_{-mt} \right) ~,
\end{equation}
where $c\left( \mathbf{d}_{mt} \right)$ and $c\left( \mathbf{d}_{-mt} \right)$ are agent $m$'s own and residual performances in period $t$, respectively (see Eq. \ref{eq:task-perf}). The parameter $a = [0,1] \in \mathbb{R}^{+} $ is the incentive parameter that defines to which extent the two performances contribute to the the agent's compensation.   

\subsection{Sub-model A: Hill climbing search}
\label{sec:hillclimbing}

Agents can use the following hill climbing algorithm to improve their utility in periods where $t\Mod\tau\neq 0$. The algorithm involves searching for solutions in the neighborhood of the previously implemented solution $\mathbf{d}_{mt-1}$ that offer higher utility. We define the neighborhood in terms of the Hamming distance of $1$. Upon finding a candidate solution $\mathbf{d}^{\ast}_{mt}$ in the neighborhood, the agent evaluates it together with the other recently implemented solutions. This allows the agent to adapt and improve its decision-making over time.

In this phase, direct communication among agents is not allowed, so agent $m$ must rely on the other agents' decisions from the previous period, $\mathbf{d}_{-mt-1}$, when evaluating potential solutions. The agent makes its decisions about which solution $\mathbf{d}_{mt}$ to implement in period $t$ according to the following rule:
\begin{equation}
\label{eq:decision-rule}
 \mathbf{d}_{mt} = 
    \begin{cases}
    \mathbf{d}_{mt-1}   &   \text{if } U(\mathbf{d}_{mt-1},\mathbf{d}_{-mt-1}) \geq U(\mathbf{d}^{\ast}_{mt},\mathbf{d}_{-mt-1})~, \\
    \mathbf{d}^{\ast}_{mt} & \text{otherwise .}
    \end{cases}
\end{equation}
The first case describes situations where the candidate solution does not offer a higher utility than the previous one, so the agent sticks with $\mathbf{d}_{mt-1}$. In the second case, the candidate solution offers a higher utility, so the agent decides to implement it in period $t$.

The solution to the entire decision problem that is implemented in period $t$ is the combination of the individual decisions made by all $M$ agents: 
\begin{equation}
\label{eq:bitstring}
\mathbf{d}_t = \left[ {\mathbf{d}_{1t}}, \dots, \mathbf{d}_{Mt}\right]~,
\end{equation}
\noindent and the performance achieved by the organization in that period is $c(\mathbf{d}_t)$ (Eq. \ref{eq:task-perf}). Therefore, the organization's performance results from the combined actions of all agents. 

\subsection{Sub-model B: Learning interdependencies}
\label{sec:learning}

Agents are aware of the potential interdependencies between decisions, but they do not know the exact structure of these interdependencies. However, agents have beliefs about them based on  observations of the consequences of their decisions in their areas of responsibility. The number of cases in which agent $m$ observed and did not observe an interdependence between decisions $d_i$ and $d_j$ up to period $t$ are stored in $\alpha_{mt}^{ij} \in \mathbb{N}$ and $\beta_{mt}^{ij} \in \mathbb{N}$, respectively. We can formalize agent $m$'s belief about the interdependencies between decisions $d_i$ and $d_j$ based on these observations using a corresponding Beta distribution:
\begin{equation}
\label{eq:beliefs}
    \mu_{mt}^{ij}= E(X)=\frac{\alpha_{mt}^{ij}}{\alpha_{mt}^{ij}+\beta_{mt}^{ij}}~,
\end{equation}
where $X \sim B(\alpha_{mt}^{ij}, \beta_{mt}^{ij})$. 

\begin{figure}
\includegraphics[width=0.9\textwidth]{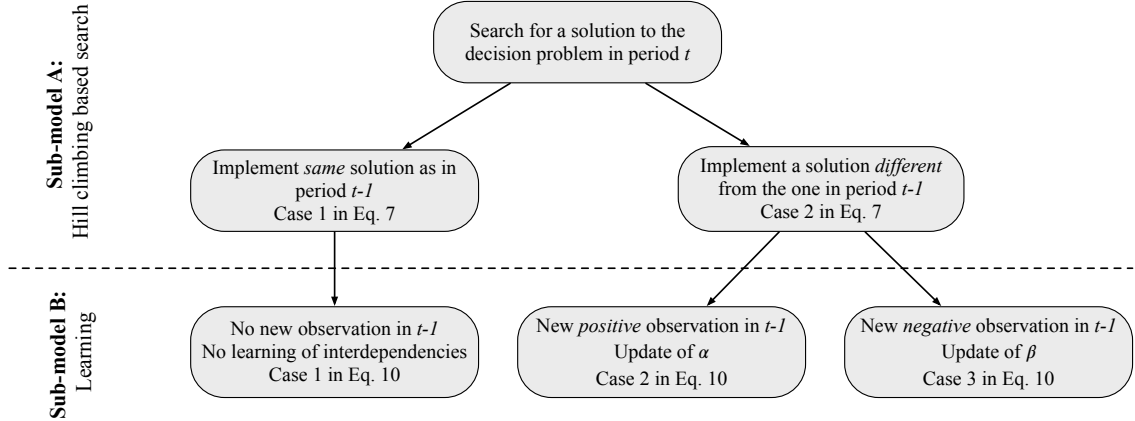}
\caption{Interrelations between sub-models A and B: Learning paths}
\label{fig:decisiontree}
\end{figure}

At the start of the simulation, all observations are set to one, which means that the initial value of $\alpha_{m0}^{ij}$ and $\beta_{m0}^{ij}$ are equal to one for all $m$, $i$, and $j$ such that $i$ and $j$ are not the same. This results in initial beliefs of $0.5$, indicating that agents initially assume that there is a $50\%$ chance of interdependencies. Then, in every period $t \Mod \tau \neq 0$, agents perform the search procedure introduced in Sec. \ref{sec:hillclimbing} and also update their beliefs in line with the following procedure: 
\begin{enumerate}
    \item Recall that the solution to agent $m$'s partial decision problem implemented in period $t$ is $\mathbf{d}_{mt}$. If the agent decides to flip a decision (i.e., the second case in Eq. \ref{eq:decision-rule}), we indicate this decision by $i$, where $d_{it} \in \mathbf{d}_{mt}$. After implementing $\mathbf{d}_{mt}$, agent $m$ observes the performance contributions $c_{jt} $ of all other decisions $d_{jt} \in \mathbf{d}_{mt}$ within their area of responsibility, with $i\neq j$.
    \item Next, agent $m$ updates the observations for all decisions $j\neq i$ in the area of responsibility as follows: 
    \begin{equation}
        \label{eq:beliefupdate}
       \left(\alpha_{mt}^{ij}, \beta_{mt}^{ij}\right)  = 
    \begin{cases}
         \left(\alpha_{mt-1}^{ij}, \beta_{mt-1}^{ij}\right)&              \text{if } \mathbf{d}_{mt}= \mathbf{d}_{mt-1}^{} ,\\
     \left(\alpha_{mt-1}^{ij}+1, \beta_{mt-1}^{ij}\right)   &   \text{if } c_{jt} \neq c_{jt-1}~ \text{ and } \mathbf{d}_{mt}= \mathbf{d}_{mt}^{\ast},\\
     \left(\alpha_{mt-1}^{ij}, \beta_{mt-1}^{ij}+1 \right)         &              \text{if } c_{jt} = c_{jt-1}~ \text{ and } \mathbf{d}_{mt}= \mathbf{d}_{mt}^{\ast},\\
    \end{cases} \\
\end{equation}
whereby $\forall i: d_{it}\in \mathbf{d}_{mt}$ and $\forall j: d_{jt}\in \mathbf{d}_{mt}:j\neq i$. Whenever agent $m$ observes a change in the performance contribution of decision $j$ from period $t-1$ to period $t$ due to a flip in decision $i$, $\alpha_{mt}^{ij}$ is increased by one (as shown in the second case in Eq. \ref{eq:beliefupdate}). Otherwise, $\beta_{mt}^{ij}$ is increased by one (as shown in the third case in Eq. \ref{eq:beliefupdate}). If agent $m$ does not flip a decision in the current period, the observations are the same as in the previous period (as shown in the first case in Eq. \ref{eq:beliefupdate}). These learning paths  are illustrated in Fig. \ref{fig:decisiontree}.
\end{enumerate}
\begin{enumerate}
    \setcounter{enumi}{2}   
    \item Finally, agents recompute their beliefs in period $t$ according to Eq. \ref{eq:beliefs}.
\end{enumerate}

Please note that agents can only observe the performance contributions \textit{within their own} areas of responsibility. If the decision problem is decomposed such that there are interdependencies with decisions from \textit{outside} an agent's area of responsibility, there may be an external influence on performance contributions that the agent cannot identify. As a result, there is the risk of learning errors as agents may \textit{wrongly} induce the existence of interdependencies from their observations.

\subsection{Sub-model C: Task allocation}
\label{sec:reallocation}

Every $\tau$ periods, agents can re-organize the allocation of tasks. During these periods, agents do not adapt the implemented solutions to their decision problems or make observations and update their beliefs about interdependencies. Instead, they exclusively focus on task-reallocation. It is important to note that re-allocating tasks can alter the agents' areas of responsibility and, thus, may affect the tasks from which they experience utility.  

To begin the process, agent $m$ offers a decision in their area of responsibility to the other agents. We will refer to the decision offered by agent $m$ as $i_m\in\{1,\dots,N\}$.\footnote{Note that every agent is required to be responsible for at least one task, so an agent can only participate in the task allocation process and offer tasks to other agents if they are currently responsible for two or more tasks.}
Next, all agents (except for the one who is offering) indicate their interest in taking over the tasks by sending signals. In the process, agents can follow two strategies: they can be myopic and focus only on immediate performance without considering long-term effects (as described in Sec. \ref{sec:utility-based-strategy}), or they can make long-sighted decisions and aim to maximize the interdependencies between the decisions in their areas of responsibility (internal interdependencies) and minimize the interdependencies with the  decision allocated to the other agents (external interdependencies) (as described in Sec. \ref{sec:interdependence-based-strategy}).
Once all signals have been received, the task is re-allocated to the agent who sent the strongest signal, and the agent who offered the decision receives a compensation payment equal to the value of the second-highest signal. 

\subsubsection{Performance-based approach}
\label{sec:utility-based-strategy}

Agents who follow this strategy are myopic decision makers, focusing only on the immediate performance contributions of the decisions they are in charge of. They offer decisions with relatively low performance contribution to other agents,, and submit signals for other decisions and hope of being allocated a task that will lead to a higher performance contribution than the amount of the corresponding compensation that they will have to pay to the offering agent. In this way, the agents' actions have an immediate effect on their utility.
The task allocation process is organized as follows:
\begin{enumerate}
    \item Agent $m$ selects the decision $i_m$ that they are willing to exchange in period $t$ and informs the other agents $r \in \{1,\dots,m-1, m+1, \dots, M\}$ about the offer. Selecting the decision $i_m$ is based on the previous period's performances, specifically it is the decison in agent $m$'s area of responsibility that is associated with the minimum performance contribution in $t-1$:
\end{enumerate}
\begin{equation}
   {i}_m \in \argmin_{ i': {d}_{it} \in \mathbf{d}_{mt-1}} c_{i't-1}~.
\end{equation}
\begin{enumerate}
    \setcounter{enumi}{1}   
    \item In addition, agent $m$ fixes a threshold $p_{i_{m}t}$ for re-allocating this decision in $t$. The threshold is the performance contribution of the offered decision, so
    \begin{equation}
    \label{eq:minperf}
        {p}_{i_{m}t}= c_{i_{m}t-1}~.
    \end{equation}
    The offered decision will only be re-allocated to another agent if the signal sent by that agent is greater than ${p}_{i_{m}t}$.
    \item Once all agents have selected the tasks they want to offer, they can submit their signals.
    However, only agents with available resources can participate in the allocation process. This means that an agent $m$ will only proceed to the next step only if $|\mathbf{d}_{mt-1}| < Q$.
    \item If agents have available resources, they will compute their signals for all offers except their own. The signals $\tilde{p}_{{i}_mt}^{r}$ submitted by an agent $r$ for a given offered decision $i_m$ is the performance contribution that the agent expects from this decision in period $t$. However, since the offered decision $i_m$ falls outside of agent $r$'s area of responsibility, they can only estimate the related performance contribution using the following formula: 
    \begin{equation}
        \label{eq:perf-basedbids}
        \tilde{p}_{{i}_mt}^{r}= c_{i_{m}t-1} + \epsilon~,
    \end{equation}
    where $\epsilon \sim N(0,\sigma)$ indicates an error term that accounts for the uncertainty in the estimate.
\end{enumerate}

\subsubsection{Interdependence-based approach}
\label{sec:interdependence-based-strategy}

If agents follow this strategy, they will not only focus on immediate performances, but they will also consider the mirroring hypothesis. This means that agents will aim to maximize the interdependencies between decisions \textit{within their own} \textit{areas of responsibility}. The task allocation process is organized as follows:
    \begin{enumerate}
        \item Agent $m$ identifies the decision $i_m$ that is being offered to the other agents in the current round using the following criteria: 
      \begin{equation}
       {i}_m \in \argmin_{i': d_{it} \in \mathbf{d}_{mt-1}}\left( \frac{1}{|\mathbf{d}_{mt-1}|-1} \sum_{\substack{j: d_{jt} \in \mathbf{d_{mt-1}} \\ j \neq i'}} \mu_{mt}^{i'j} \right)
    \end{equation}  
    As a reminder, agents want to maximize internal and minimize external interdependencies in this strategy. Equation \ref{eq:mink} returns the decision that is associated with the minimum average belief about interdependencies between decision $i_m$ and the other decisions in agent $m$'s area of responsibility. 
    \item Again, agent $m$ will fix a threshold $p_{i_{m}t}$ for re-allocating decision $i_m$ to other agents in period $t$. For simplicity, the average belief about internal interdependencies is used as the threshold value:
    \begin{equation}
      \label{eq:mink}
        p_{i_{m}t} = \frac{1}{|\mathbf{d}_{mt-1}|-1} \sum_{\substack{j: d_{jt} \in \mathbf{d_{mt-1}} \\ j \neq i_m}} \mu_{mt}^{i{_m}j} 
    \end{equation}
    \item Once all agents prepared their offers, they cab proceed to compute and send their signals. However, they will only move on to the next step if they have sufficient resources, i.e., if $|\mathbf{d}_{mt-1}| < Q$.
    \item In period $t$, agents $r \in \{1,\dots,m-1, m+1, \dots, M\}$ send a signal containing the average belief about the interdependencies between the offered decision $i_m$ and the decisions within their areas of responsibility $\mathbf{d}_{rt-1}$. Agent $r$'s signal for decision $i_m$ in period $t$ is computed according to:
    \begin{equation}
        \label{eq:k-basedbids}
        \tilde{p}_{i_{m}t}^{r}= \frac{1}{|\mathbf{d}_{rt}|} \sum_{{ j: d_{jt} \in \mathbf{d_{rt}}}} \mu_{rt}^{i{_m}j}
    \end{equation}
\end{enumerate}

\subsubsection{Task allocation}
Once all agents sent their signal, there are exactly $M-1$ signals for each offer $i_m$ . Recall that agent $m$ offered decision $i_m$ at a threshold signal of ${p}_{i_{m}t}$ and the other agents sent their signals $\tilde{p}^{r}_{i_{m}t}$.
We can denote the set of signals received for decision $i_m$ in period $t$ by the vector $\mathbf{P}_{i_{}t}$, and we can compute
the maximum signal for decision $i_m$ in period $t$ by 
${p}^{r\ast}_{i_{m}t}= \max_{p'\in \mathbf{P}_{i_{}t}} (p')$. The agent who sends this signal is denoted by $r^{\ast}$. 
The tasks are (re-)allocated as follows
\begin{enumerate}
    \item If the the maximum signal ${p}^{r\ast}_{i_{m}t}$ is equal to or exceeds the threshold ${p}^{}_{i_{m}t}$, the decision $i_m$ is re-allocated from agent $m$ to agent $r^{\ast}$ according to
    \begin{subequations}
    \begin{eqnarray}
      \mathbf{d}_{mt} & = & \mathbf{d}_{i_{m}t-1} \setminus \{ d_{i_{m}t-1} \} ~\text{and}\\
        \mathbf{d}_{r^{\ast}t} & = & \left[ {\mathbf{d}_{r^{\ast}t-1}},d_{i_{m}t-1}  \right]~,
        \end{eqnarray}
    \end{subequations}
    \noindent where $\setminus$ indicates the complement. If the second highest signal exceeds (does not exceed) the threshold, agent $r^{\ast}$ gets charged the second highest bid (threshold). 
    \item If the the maximum signal ${p}^{r\ast}_{i_{m}t}$ does not exceed the threshold ${p}^{}_{i_{m}t}$, agent $m$ remains responsible for decision $i_m$, so  
    \begin{equation}
        \mathbf{d}_{mt}:=\mathbf{d}_{mt-1}~.
    \end{equation}
    \item Finally, agents do not update their beliefs about interdependencies in periods in which task are re-allocated. Therefore, the observations are the same as in the previous period, i.e.,  $(\alpha^{ij}_{mt}, \beta^{ij}_{mt}) = (\alpha^{ij}_{mt-1}, \beta^{ij}_{mt-1})$.
\end{enumerate}

\paragraph{Comparison of the two approaches} 
The two approaches for determining offers and computing signals define the agents' behaviors in different ways. The performance-based strategy focuses on increasing immediate performance contributions without considering long-term effects. Agents using this strategy offer tasks in their area of responsibility that are associated with the minimum performance, and they hope for compensation payments that are higher than the utility they would receive from performing the tasks on their own. This behavior aligns with the concept of myopic utility maximization \citep{simon1967}.
In contrast, the interdependence-based strategy does not prioritize short-term utility. Instead, it follows the concept of the mirroring hypothesis, which means that agents aim to minimize interdependencies between their own and other agents' areas of responsibility \citep{colfer2016}. They hope that this will lead to greater control and higher utility in the long run. Overall, the choice of strategy affects the way that agents behave and the decisions they make.

\subsection{Simulation setup and observations}

\begin{figure}
     \centering
      \vspace{-4mm}
     \begin{subfigure}[b]{0.33\textwidth}
         \centering
             \caption{Small diagonal blocks ($K=2$)}
         \includegraphics[width=\textwidth]{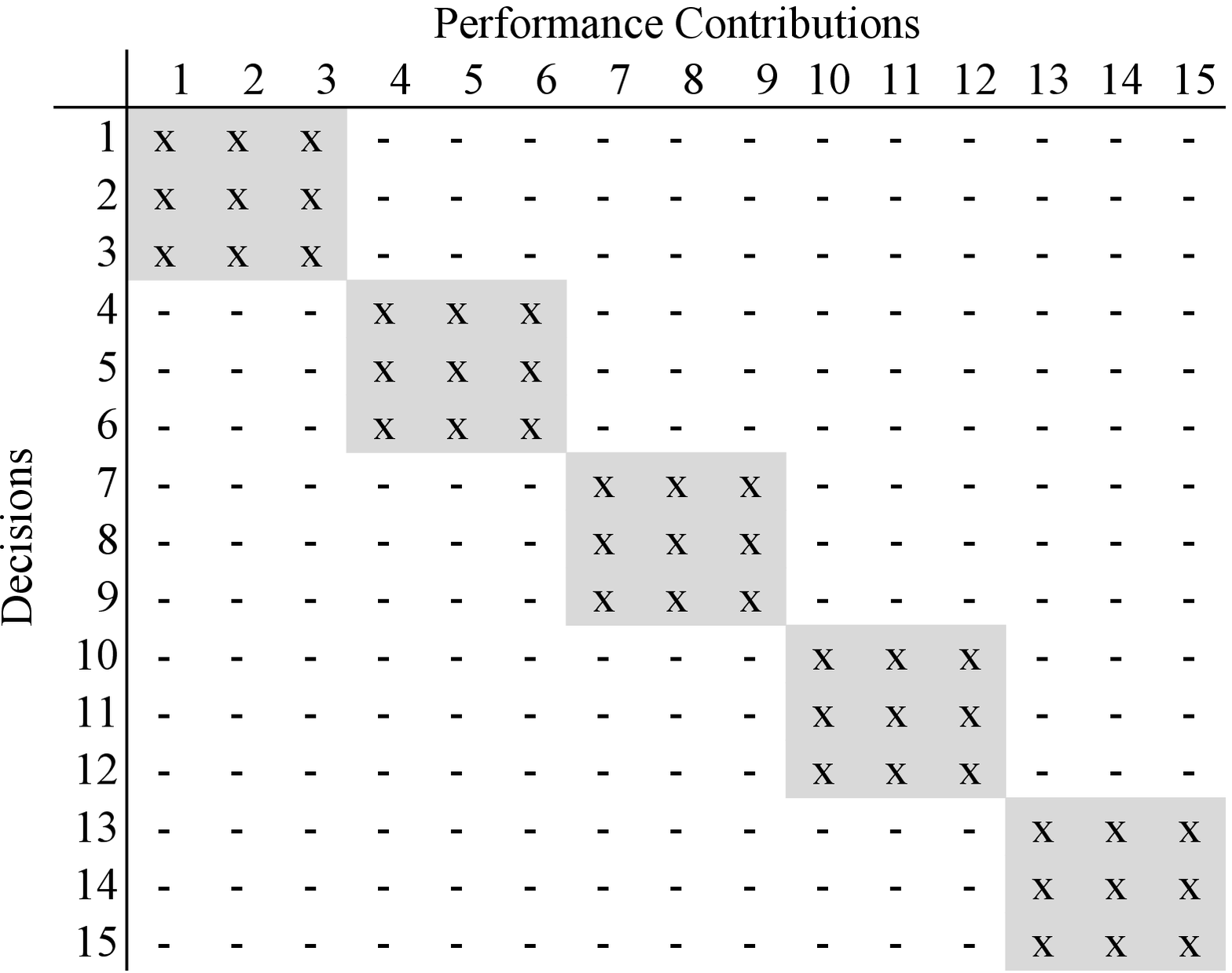}
         \label{fig:matrix-5mal3}
     \end{subfigure}
      \vspace{-4mm}
      \hspace{10mm}
     \vspace{-4mm}
    \begin{subfigure}[b]{0.33\textwidth}
         \centering
                  \caption{Big diagonal blocks ($K=4$)}
         \includegraphics[width=\textwidth]{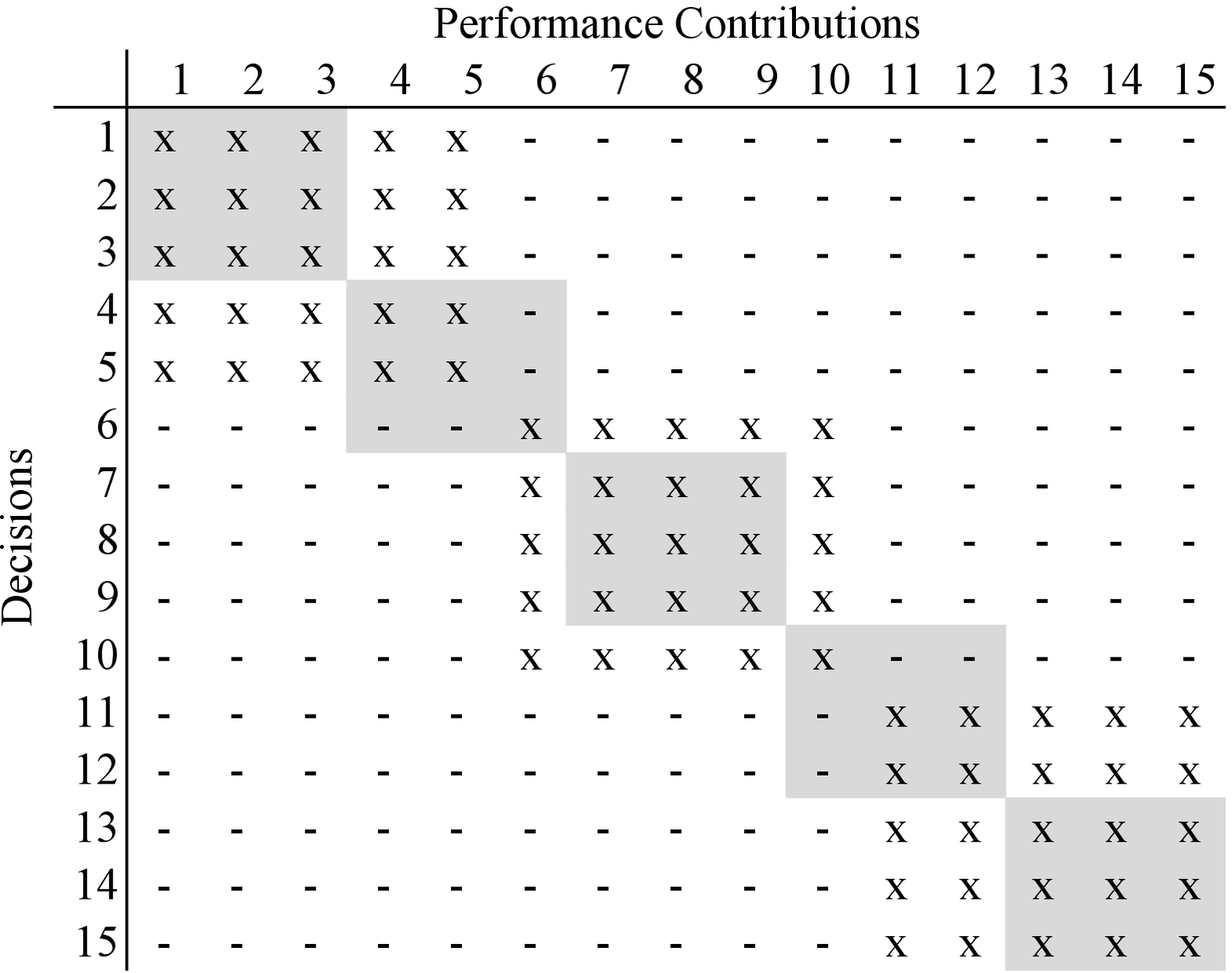}
         \label{fig:matrix-3mal5}
     \end{subfigure}
     \vspace{-4mm}
    \begin{subfigure}[b]{0.33\textwidth}
         \centering
              \caption{Small blocks: Reciprocal ($K=6$)}
         \includegraphics[width=\textwidth]{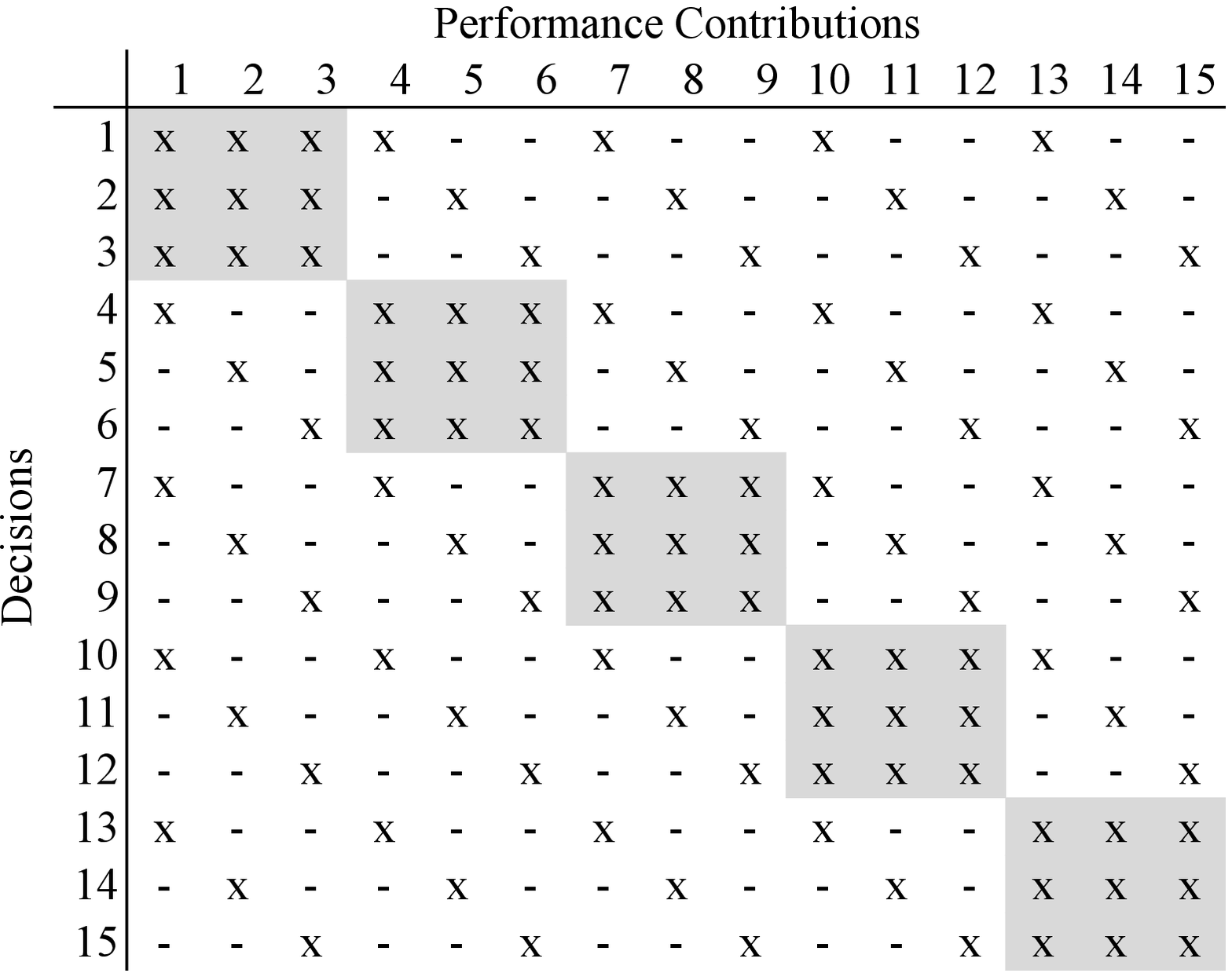}
         \label{fig:matrix-5mal3reciprocal}
     \end{subfigure}
      \vspace{-4mm}
       \hspace{10mm}
    \begin{subfigure}[b]{0.33\textwidth}
         \centering
                  \caption{Big blocks: Reciprocal ($K=6$)}
         \includegraphics[width=\textwidth]{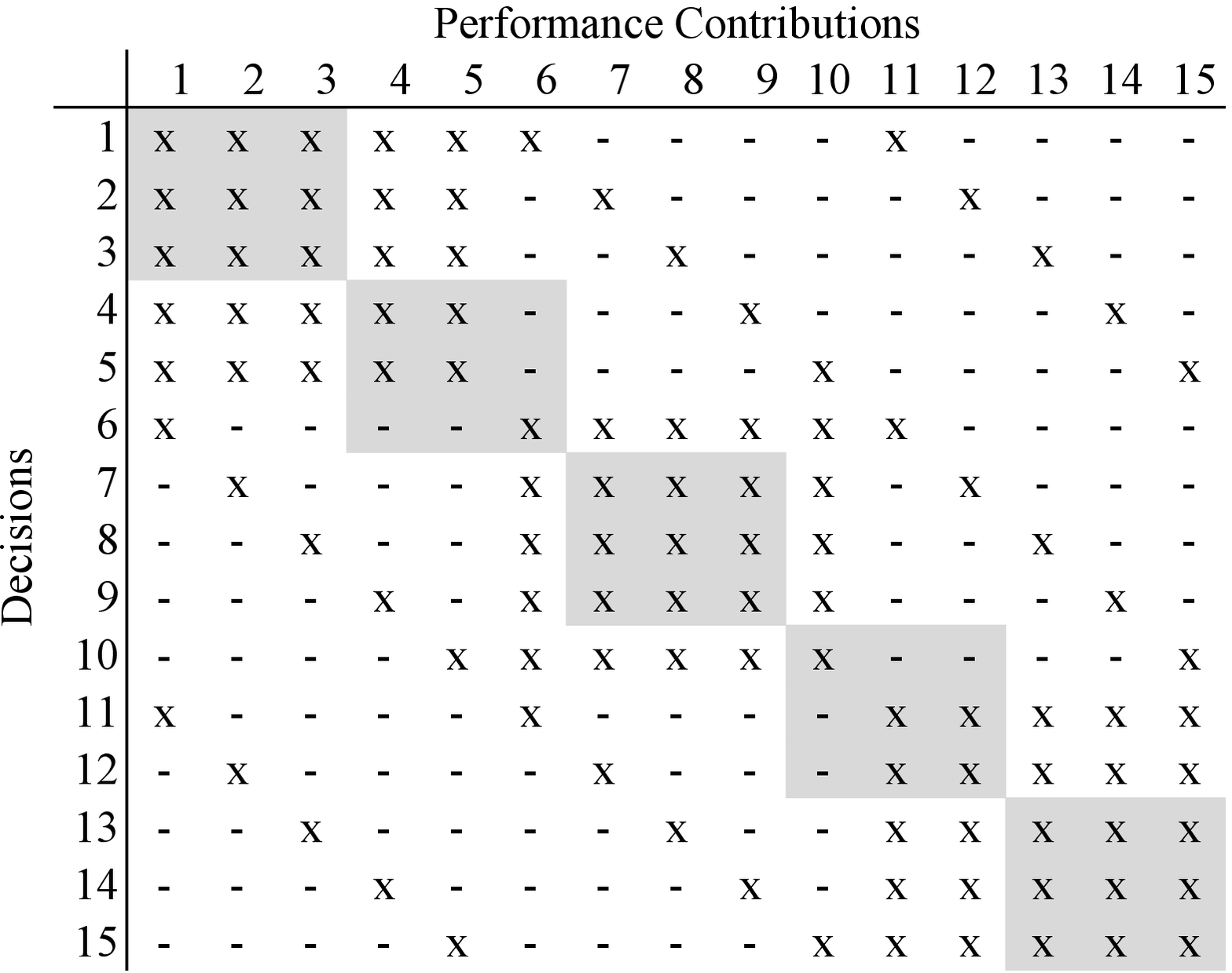}
         \label{fig:matrix-3mal5reciprocal}
     \end{subfigure}
      \vspace{-4mm}
    \begin{subfigure}[b]{0.33\textwidth}
         \centering
                  \caption{Small blocks: Ring ($K=5$)}
         \includegraphics[width=\textwidth]{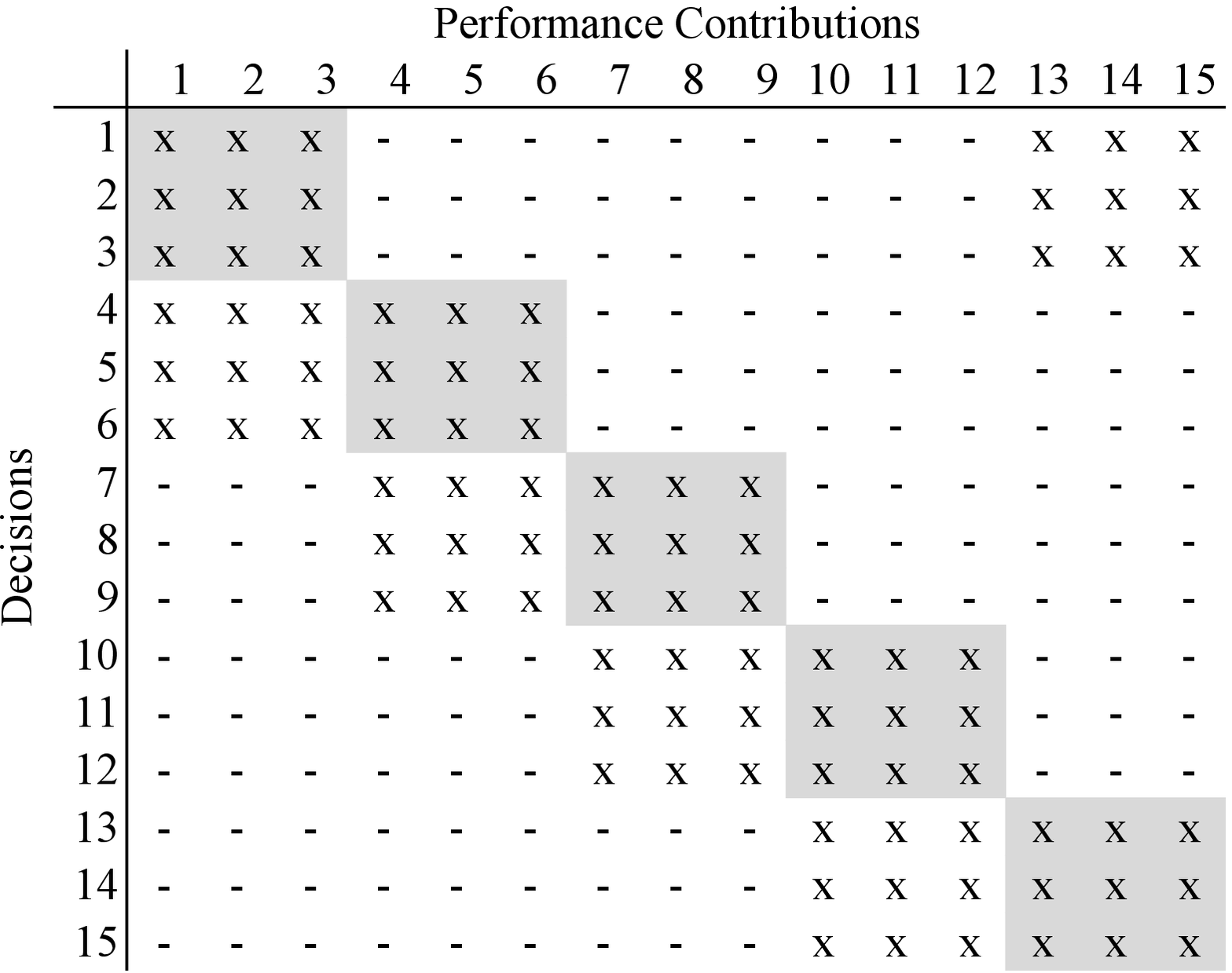}
         \label{fig:matrix-5mal3ring}
     \end{subfigure}
      \vspace{-4mm}
       \hspace{10mm}
    \begin{subfigure}[b]{0.33\textwidth}
         \centering
                  \caption{Big blocks: Ring ($K=9$)}
         \includegraphics[width=\textwidth]{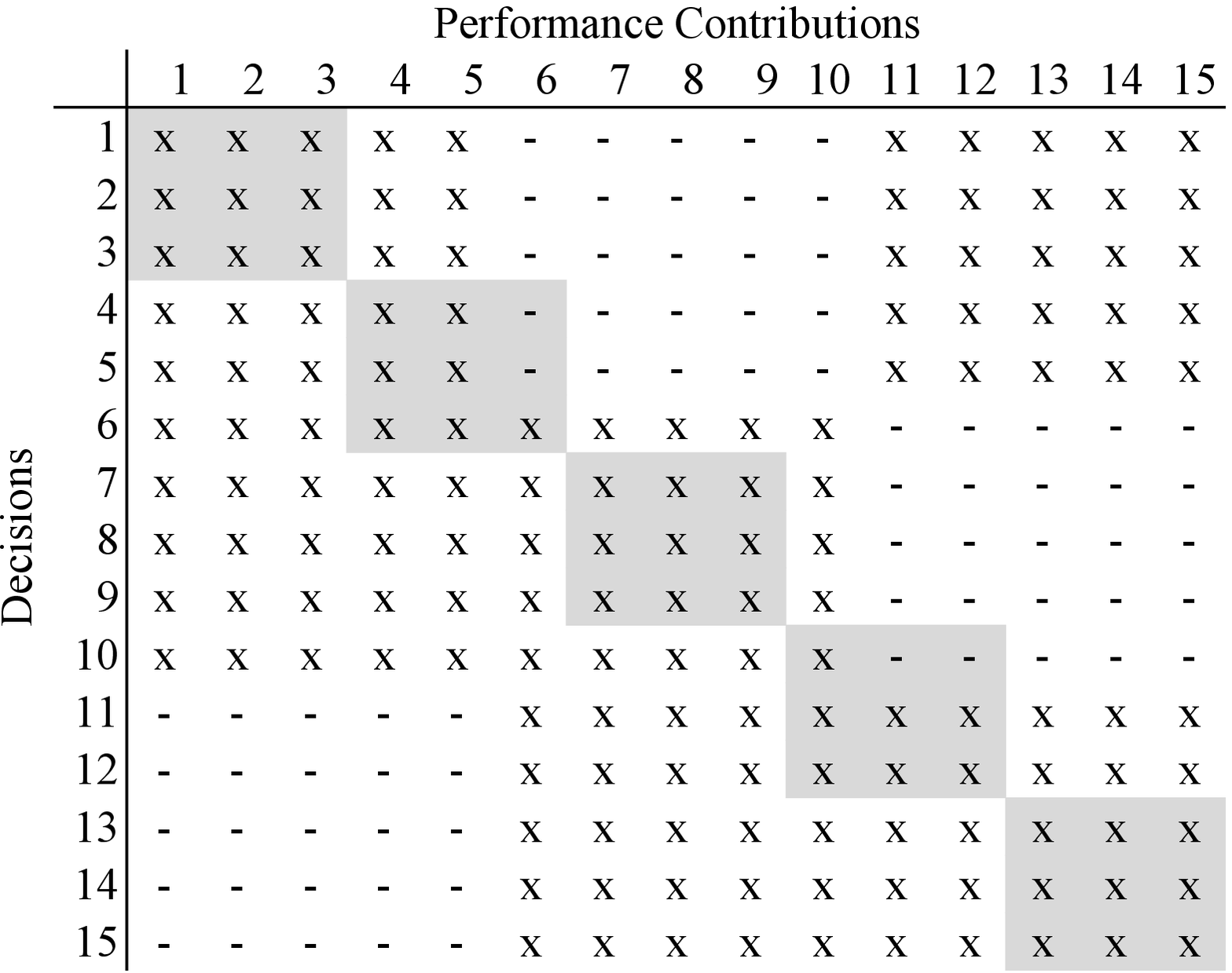}
         \label{fig:matrix-3mal5ring}
     \end{subfigure}
      \vspace{-4mm}
    \begin{subfigure}[b]{0.33\textwidth}
         \centering
                  \caption{Random pattern ($K=4$)}
         \includegraphics[width=\textwidth]{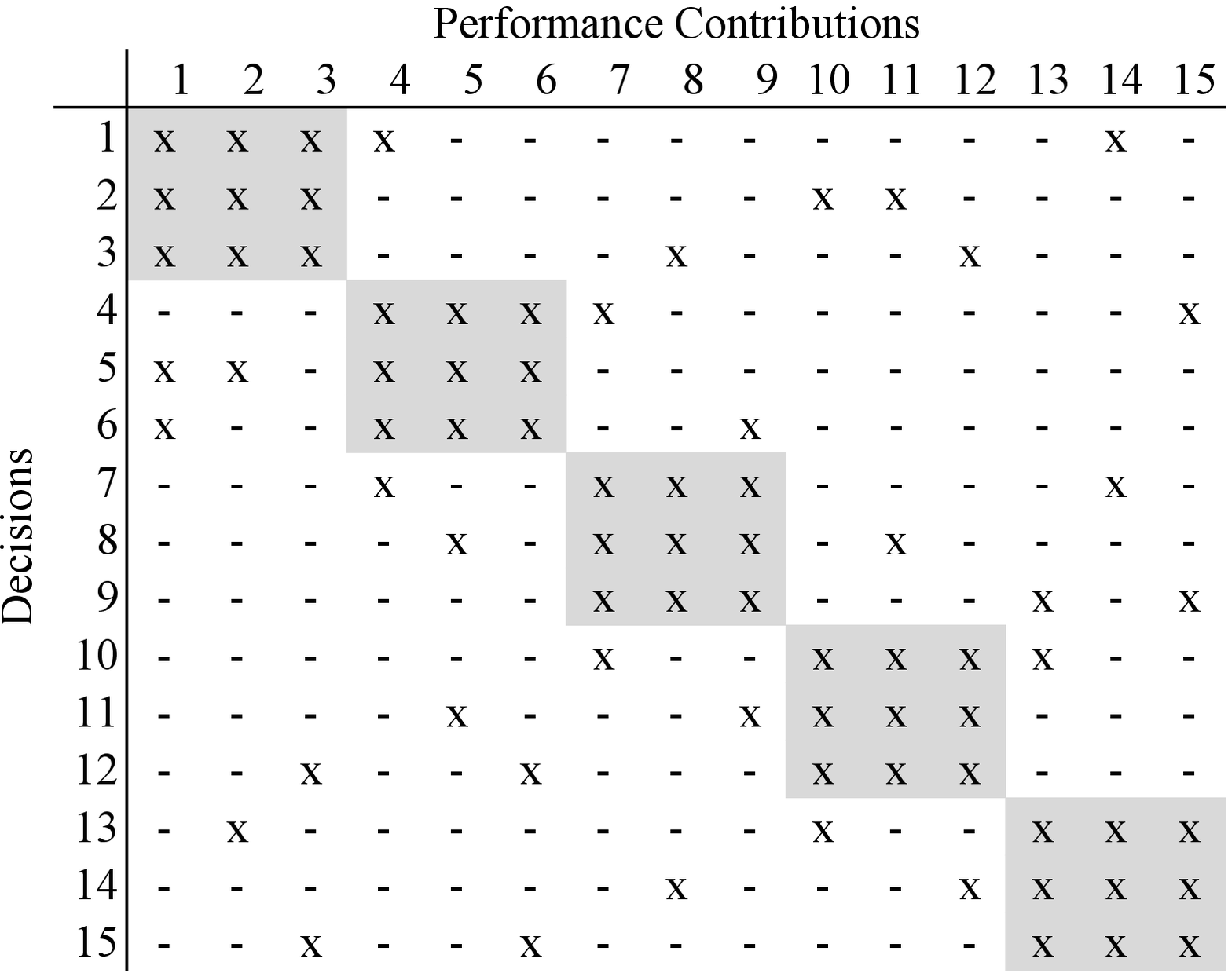}
         \label{fig:matrix-randomk4}
     \end{subfigure}
      \vspace{-4mm}
       \hspace{10mm}
    \begin{subfigure}[b]{0.33\textwidth}
         \centering
                  \caption{Random pattern ($K=6$)}
         \includegraphics[width=\textwidth]{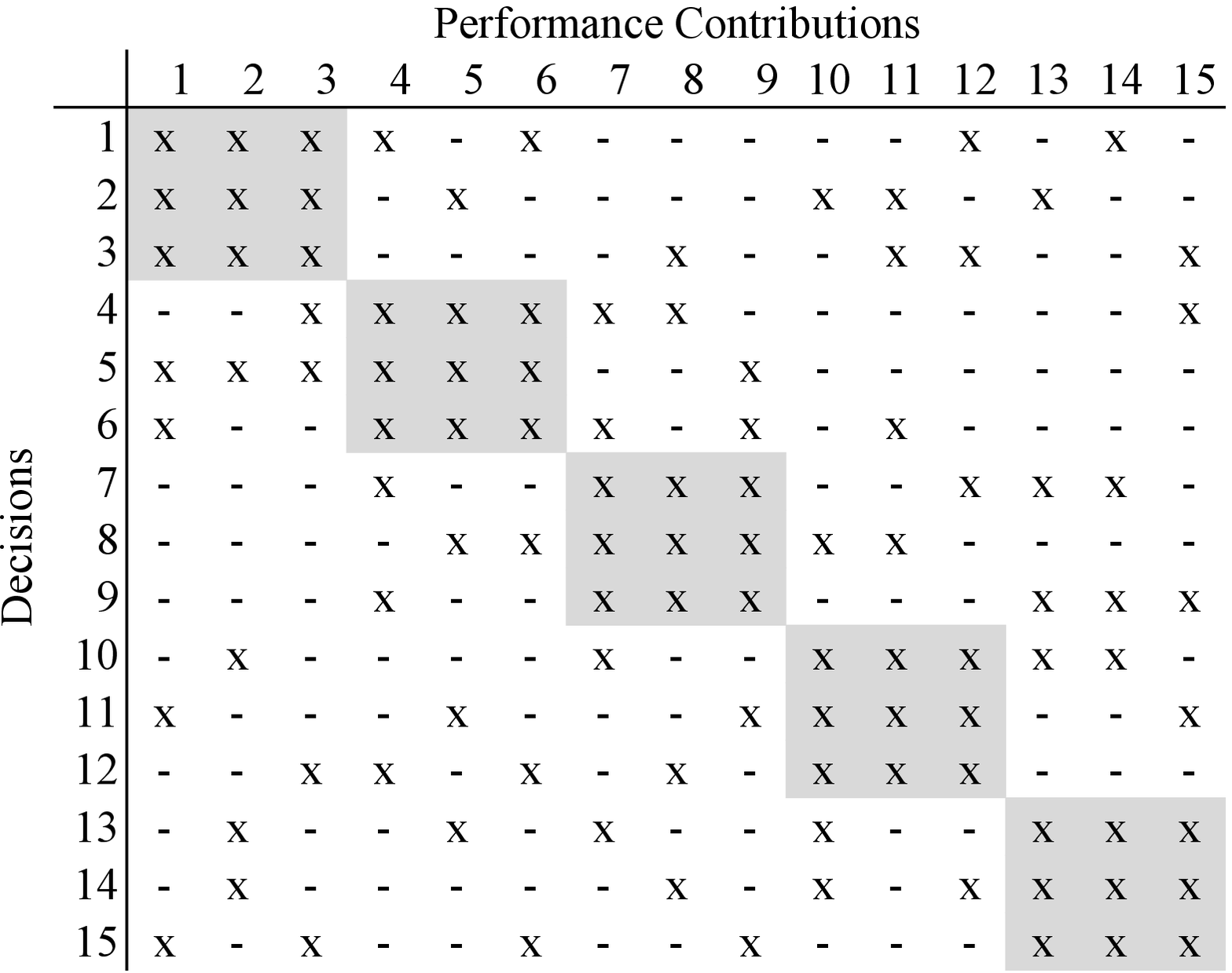}
         \label{fig:matrix-randomk6}
     \end{subfigure}

    \caption{Interdependence patterns}
    \label{fig:matrices}
\end{figure}

This paper focuses on the analysis of the effects of four main parameters on the performance and emergent task allocation. These parameters are:
\begin{enumerate}[label=(\roman*)]
\item The information used in the task allocation process, specifically the performance-based and interdependence-based strategies introduced in earlier sections of the paper.
\item The parameter $a$ of the linear incentive scheme introduced in Eq. \ref{eq:utility}, which ranges from very altruistic to very individualistic incentives. In particular, values for $a$ between $0.05$ and $1$ in steps of $0.05$ are considered.
\item The frequency of the task allocation process indicated by parameter $\tau$. The paper considers the values of $5, 15, 25$, and $35$. In addition, benchmark scenarios in which the task allocation is exogenously fixed and cannot be changed thereafter. For the benchmark cases, $\tau=\infty$. 
\item The eight patterns of interactions between tasks included in Fig. \ref{fig:matrices}, which feature big and small diagonal blocks along the main diagonal (Figs. \ref{fig:matrix-5mal3} and \ref{fig:matrix-3mal5}), reciprocal interdependencies between these blocks (Figs. \ref{fig:matrix-5mal3reciprocal} and \ref{fig:matrix-3mal5reciprocal}), and ring-like interdependencies (Figs. \ref{fig:matrix-5mal3ring} and \ref{fig:matrix-3mal5ring}). In addition, Figs. \ref{fig:matrix-randomk4} and \ref{fig:matrix-randomk6} add random interdependencies to the pattern of small diagonal blocks. The benchmark task allocation is indicated by shaded areas and reflects a sequential and symmetric task allocation. This means that agent 1 is responsible for tasks 1 to 3, agent 2 is responsible for tasks 4 to 6, and so forth. Thus, the task allocation in the benchmark cases reflects the interdependencies included in Fig. \ref{fig:matrix-5mal3}.
\end{enumerate}

\begin{table}[htp]
\begin{scriptsize}
\caption{Simulation parameters}
\label{tab:variables}
\begin{tabular}{llll}
\multicolumn{1}{l}{Type}            & \multicolumn{1}{l}{Name} & Notation        & Values \\ 
\noalign{\smallskip} 
\hline 
\noalign{\smallskip} 
\multirow{3}{*}{Independent variables}
    & Basis for computing offers/signals      &   ---           & Performance, interdependencies \\
    & Incentive parameter   &   $a$         & $\{0.05:0.05:1\}$\\
    & Periods between task re-allocation    &  $\tau$                   & $\{5,15,25,35,\infty\}$ \\
    & Task complexity       &   ---         & see Fig. \ref{fig:matrices}\\
\noalign{\smallskip} \hline \noalign{\smallskip}  
Dependent variables
    & Task performance    & $\tilde{c}(\mathbf{d}_{ts})$           & $\left[0,1\right]$\\ 
    & Task allocation    &  $\mathbf{d}_{mts}$  & Vectors of decisions \\
& Number of re-allocated tasks & $\mathbf{d}^{\text{swap}}_t$ & $\{1,\dots,5\}$\\
\noalign{\smallskip} \hline \noalign{\smallskip} 
\multirow{7}{*}{Other parameters}   
    & Time steps            & $t$                   & $\{1,\dots,200\}$\\
    & Observation period    & $T$                   & $200$ \\ 
    & Number of decisions   & $N$                   & $15$ \\ 
    & Number of agents      & $M$                   & $5$ \\ 
    & Cognitive capacity    & $Q$                   & $5$ \\ 
    & Number of simulations & $S$                   & $800$ \\ 
\noalign{\smallskip} \hline
\end{tabular}
\end{scriptsize}
\end{table}

During the simulation experiments, three variables are subject to observation: task performance, the emergent task allocation, and the number of tasks re-allocated. The performance of the solution to the entire decision problem is observed in every period $t$ and simulation run $s \in \{1,\dots,S\} \subset \mathbb{N}$ and denoted by $\mathbf{d}_{ts}$ (see Eq. \ref{eq:bitstring}). The performance of that solution is then measured by $c(\mathbf{d}_{ts})$ (see Eq. \ref{eq:task-perf}). To compare the performance across simulation runs, we normalize the observed performance $c(\mathbf{d}_{ts})$ by the maximum attainable performance on that landscape, $c(\mathbf{d}^{\ast}_s)$, using the formula:
\begin{equation}
\tilde{c}(\mathbf{d}_{ts})=\frac{c(\mathbf{d}_{ts})}{c(\mathbf{d}^{\ast}_s)}~.
\end{equation}
In addition to the performance, we also observe the allocation of tasks to agents at every period. Specifically, we record the areas of responsibility $\mathbf{d}_{mts}$ of all agents in all periods and simulation runs. It is important to note that the performance is measured based on the solution to the entire decision problem, whereas task allocation is based on the decisions included in the agents' sub-problems. Therefore, the first observation provides a performance perspective, while the second observations offers insights into the emergent organizational structures. Finally, to understand the dynamics of emergent task allocation, we also track the number of re-allocated tasks during each simulation run. 

\subsection{Data analysis}
\label{sec:data-analysis}

To analyze the \textit{functional dependencies between the dependent and the independent variables} included in Tab. \ref{tab:variables}, we use regression neural networks and compute partial dependencies. This approach is consistent with the data analysis methods suggested in \cite{patel2018}, \cite{law2015}, and \cite{blanco2022}, which recommend using regression analyses to study parameter importance and understand the emergence of patterns. 
Let $\mathbf{X}$ be the set of all independent variables included in Tab. \ref{tab:variables}. The subset $\mathbf{X}^s$ includes the independent variable(s) that are in the scope of the analysis, while $\mathbf{X}^c$ consists of the complementary set of $\mathbf{X}^s$ in $\mathbf{X}$. Then, $f(\mathbf{X})=f(\mathbf{X}^s,\mathbf{X}^c)$ represents the trained regression model. The partial dependence of the performance on the independent variables in scope is defined by the expectation of the performance with respect to the complementary independent variables, as follows: 
\begin{equation}
\label{eq:dependencies}
    f^s(\mathbf{X}^s)= E_c(f(\mathbf{X}^s,\mathbf{X}^c)) \approx \frac{1}{V}\sum_{i=1}^{V} f(\mathbf{X}^s,\mathbf{X}_{(i)}^c)~,
\end{equation}
\noindent where $V$ is the number of independent variables in $\mathbf{X}^c$ and $\mathbf{X}_{(i)}^c$ is the $i^{th}$ element. By marginalizing over the independent variables in $\mathbf{X}^c$, we obtain a function that depends only on the independent variables in $\mathbf{X}^s$.

To study the \textit{modularity of the emergent task allocation}, we employ the following metric \citep{leitner2022}: We already know that agent $m$'s decision problem in period $t$ and simulation run $s$ covers the decisions included in $\mathbf{d}_{mts}$, and the parameter $K$ describes the interdependencies of a particular decision and all performance contributions. Let $K^{\text{int}}_{mts}$ be the number of interdependencies \textit{within} agent $m$'s sub-problem in period $t$ and simulation run $s$, and $K^{\text{all}}_{mts}=|\mathbf{d}_{mts}|\cdot K$ be the number of all interdependencies between the decisions in agent $m$'s area of responsibility and all performance contributions.\footnote{Recall that $|\cdot|$ returns the length of a vector. Therefore, $|\mathbf{d}_{mts}|$ is the number of decision for which agent $m$ is responsible in period $t$ and simulation run $s$.} The modularity metric is then defined as the ratio of interdependencies within agent $m$'s decision problem ($K^{\text{int}}_{mts}$, numerator) to the total number of times the decisions assigned to agent $m$ affect all performance contributions ($K^{\text{all}}_{mts}$, denominator):

\begin{equation}
\label{eq:ratio}
    \text{Mod}_{mts} = \frac{K^{\text{int}}_{mts}}{K^{\text{all}}_{mts}}~.
\end{equation}

To illustrate how the modularity metric works, consider the case where we focus on agent 1 and the task allocation follows the benchmark case of a symmetric and sequential allocation, as shown by the shaded areas in Fig. \ref{fig:matrices}. In this scenario, agent 1 is responsible for decisions 1 to 3. For small diagonal blocks (Fig. \ref{fig:matrix-5mal3}), the number of internal interdependencies for agent 1 is $K^{\text{int}}_{1ts}=6$ and the total number of interdependencies for the decisions allocated to agent 1 is also $K^{\text{all}}_{^ts}=6$. In this case, the modularity of the benchmark solution is $\text{Mod}_{mts}=1$. If we move to small blocks and reciprocal patterns (Fig. \ref{fig:matrix-3mal5reciprocal}), the number of internal interdependencies remains $K^{\text{int}}_{1ts}=6$. However, due to a more complex decision problem, the total number of interdependencies increases $K^{\text{all}}_{^ts}=18$. As a result, the modularity of the benchmark solution becomes $\text{Mod}_{mts}=0.3\dot{3}$. It is important to note that for the analysis, the task allocation that emerges from the agents' decisions (and not the benchmark task allocation) is used to compute the respective modularity, and the goal is to compare the modularity of the emergent solution with that of the benchmark solution. 
A review of related modularity metrics can be found in \cite{peng2018}. They identify three desired features of such metrics\footnote{Please note that the context of their analysis included product modularity and this paper focuses on tasks rather than products. Therefore, whenever \cite{peng2018} refer to components and functions, this refers to tasks and performance contributions in the context of this paper}: 
\textit{(i)} Does the metric consider modularity at the levels of the task and the organization? 
\textit{(ii)} Are interdependencies between tasks and performance contributions considered? 
\textit{(iii)} Does organizational modularity consider task assignment to agents?
The task allocation metric used in this paper satisfies these requirements because it measures the actual task allocation (requirement \textit{(iii)}) against the benchmark task allocation derived from interdependencies between tasks (requirements \textit{(i)} and \textit{(ii)}). 

\section{Results and discussion}
\label{sec:results}

\subsection{The effects of incentives in bottom-up task allocation}
\label{sec:results-performance}

\begin{figure}
     \centering
     \begin{subfigure}[b]{0.33\textwidth}
         \centering
                  \caption{Small diagonal blocks ($K=2$)}
         \includegraphics[width=\textwidth]{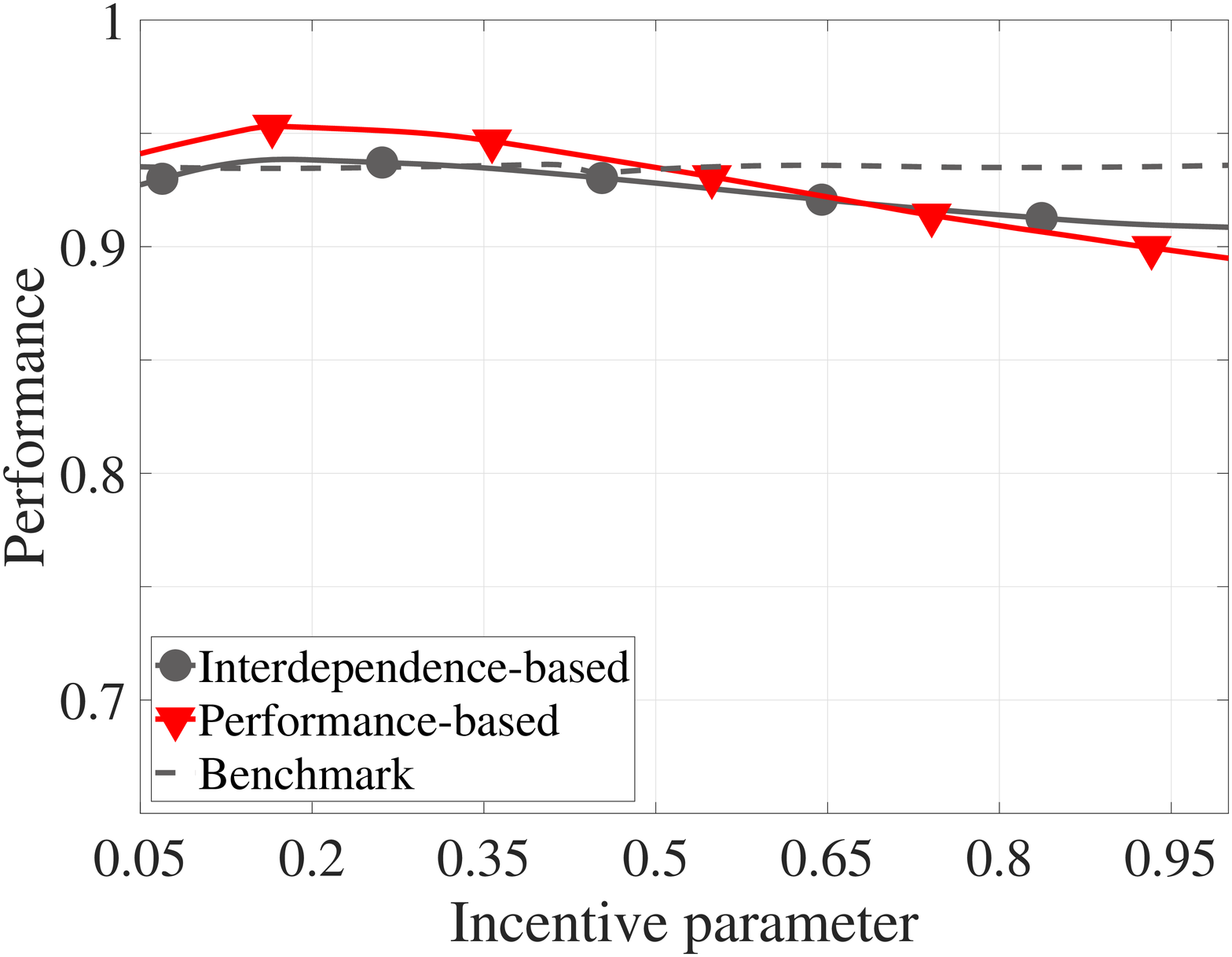}
         \label{fig:alpha-5mal3}
     \end{subfigure}
       \vspace{-4mm}
        \hspace{10mm}
    \begin{subfigure}[b]{0.33\textwidth}
         \centering
                  \caption{Big diagonal blocks ($K=4$)}
         \includegraphics[width=\textwidth]{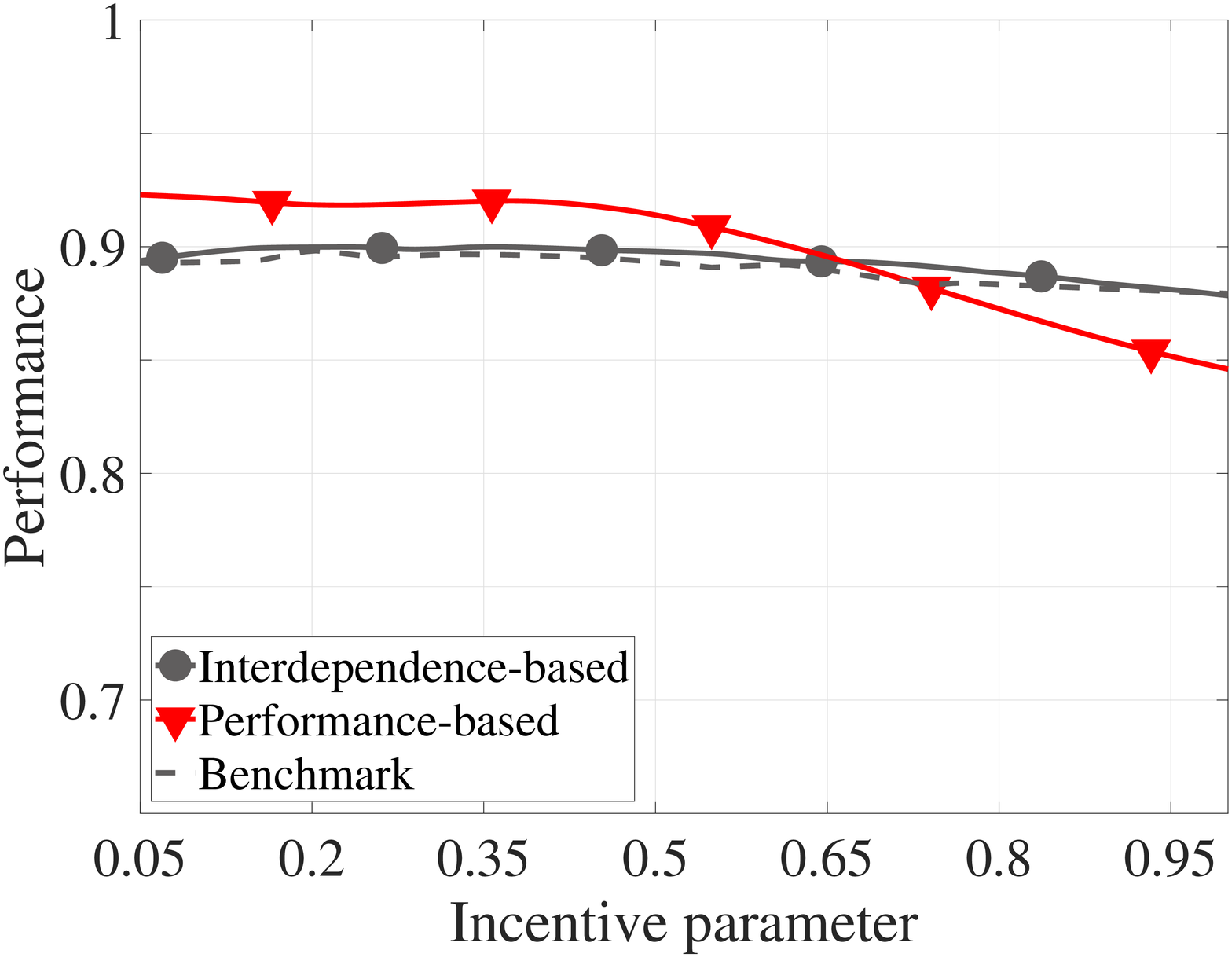}
         \label{fig:alpha-3mal5}
     \end{subfigure}
      \vspace{-4mm}
      
    \begin{subfigure}[b]{0.33\textwidth}
         \centering
                  \caption{Small blocks: Reciprocal ($K=6$)}
         \includegraphics[width=\textwidth]{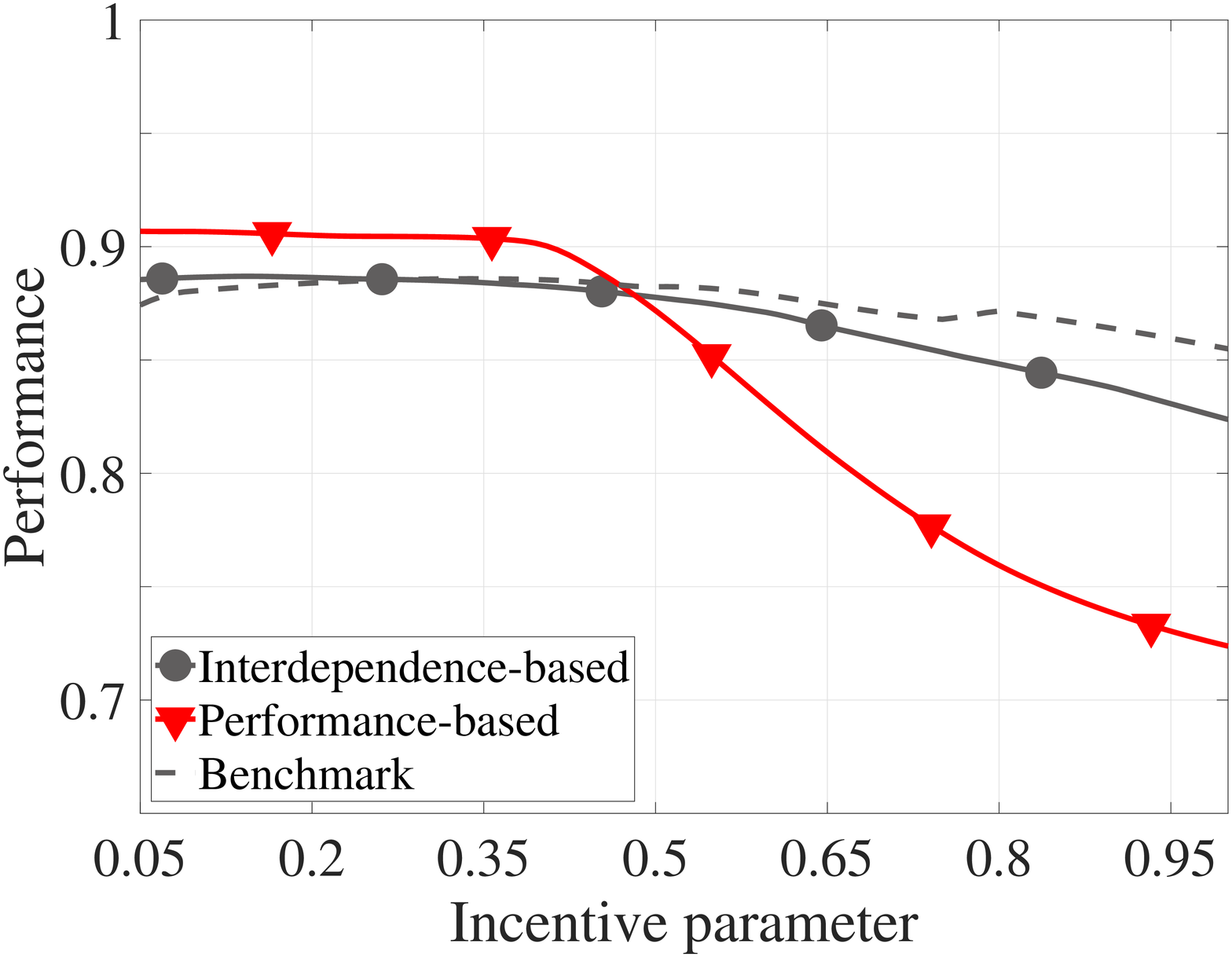}
         \label{fig:alpha-5mal3reciprocal}
     \end{subfigure}
       \vspace{-4mm}
        \hspace{10mm}
    \begin{subfigure}[b]{0.33\textwidth}
         \centering
                  \caption{Big blocks: Reciprocal ($K=6$)}
         \includegraphics[width=\textwidth]{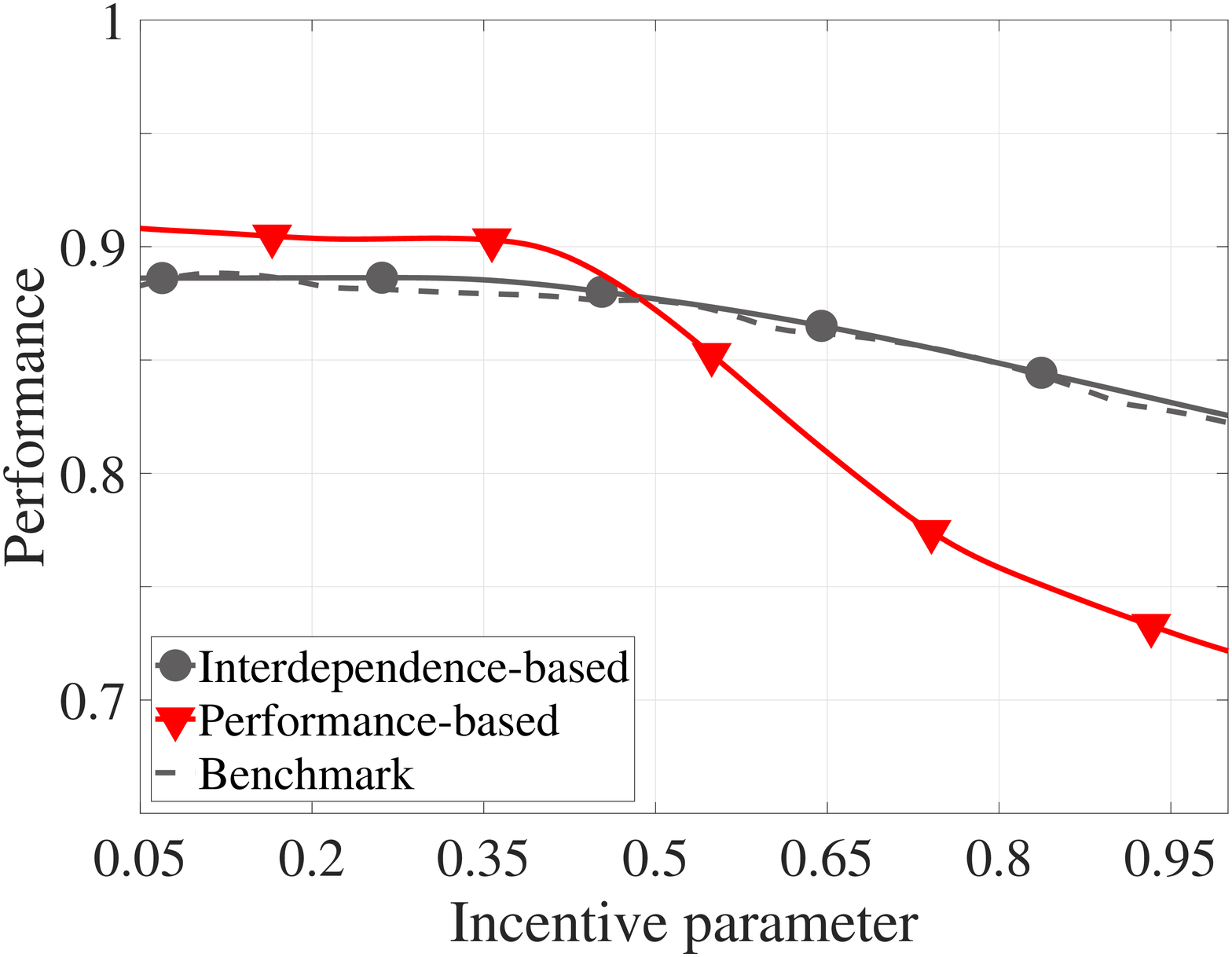}
         \label{fig:alpha-3mal5reciprocal}
     \end{subfigure}
       \vspace{-4mm}
     
    \begin{subfigure}[b]{0.33\textwidth}
         \centering
                  \caption{Small blocks: Ring ($K=5$)}
         \includegraphics[width=\textwidth]{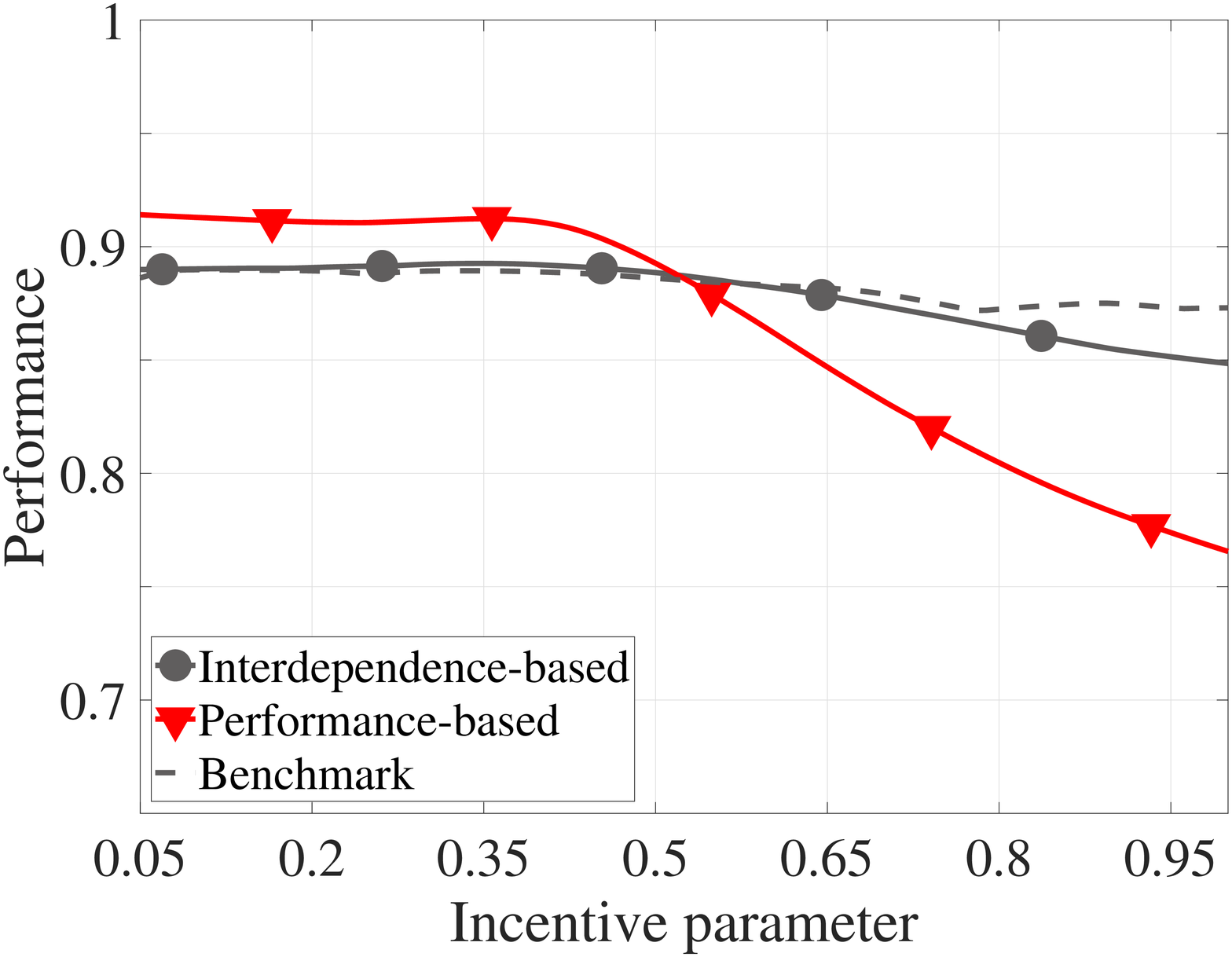}
         \label{fig:alpha-5mal3ring}
     \end{subfigure}
       \vspace{-4mm}
        \hspace{10mm}
    \begin{subfigure}[b]{0.33\textwidth}
         \centering
                  \caption{Big blocks: Ring ($K=9$)}
         \includegraphics[width=\textwidth]{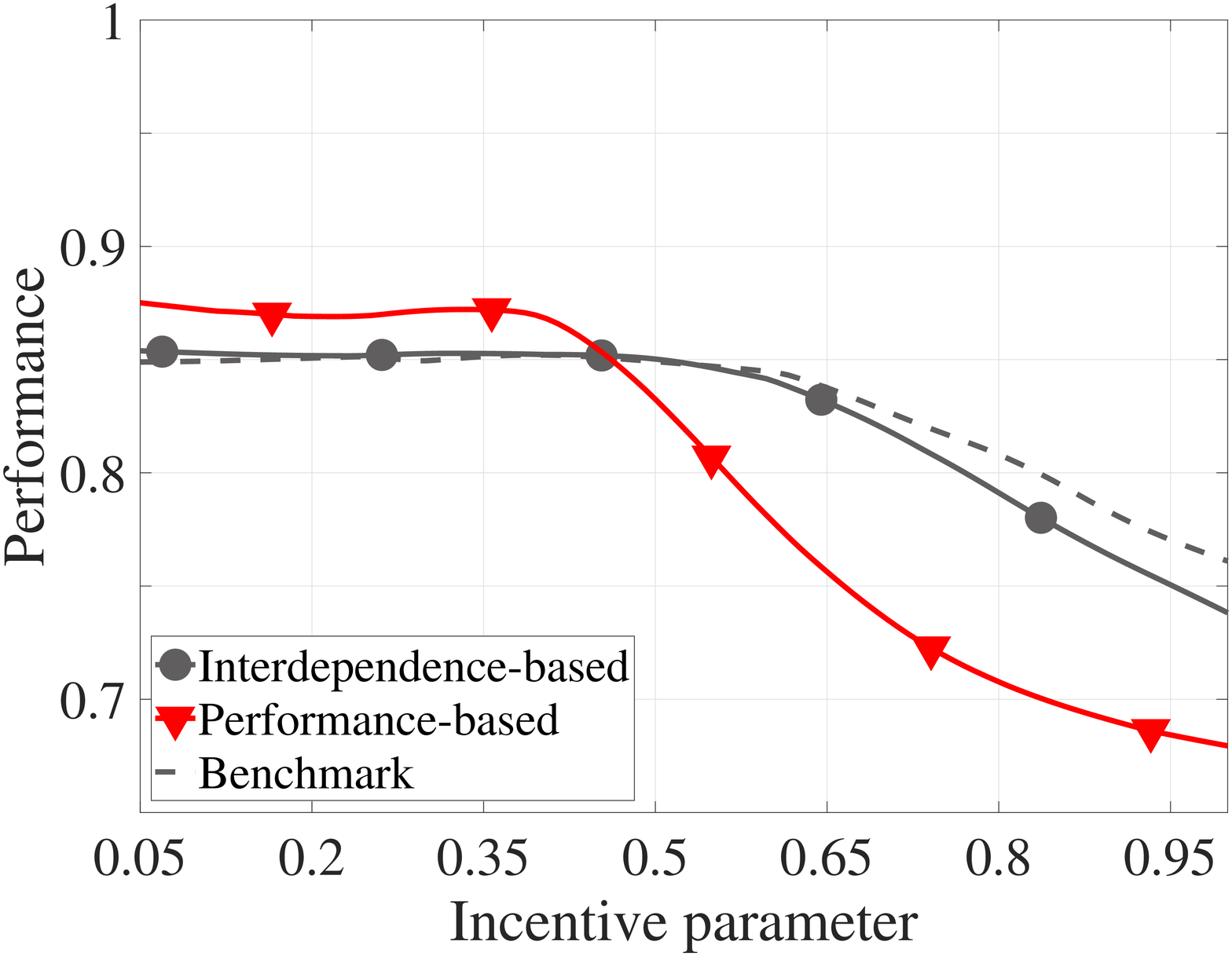}
         \label{fig:alpha-3mal5ring}
     \end{subfigure}
       \vspace{-4mm}
     
    \begin{subfigure}[b]{0.33\textwidth}
         \centering
                  \caption{Random pattern ($K=4$)}
         \includegraphics[width=\textwidth]{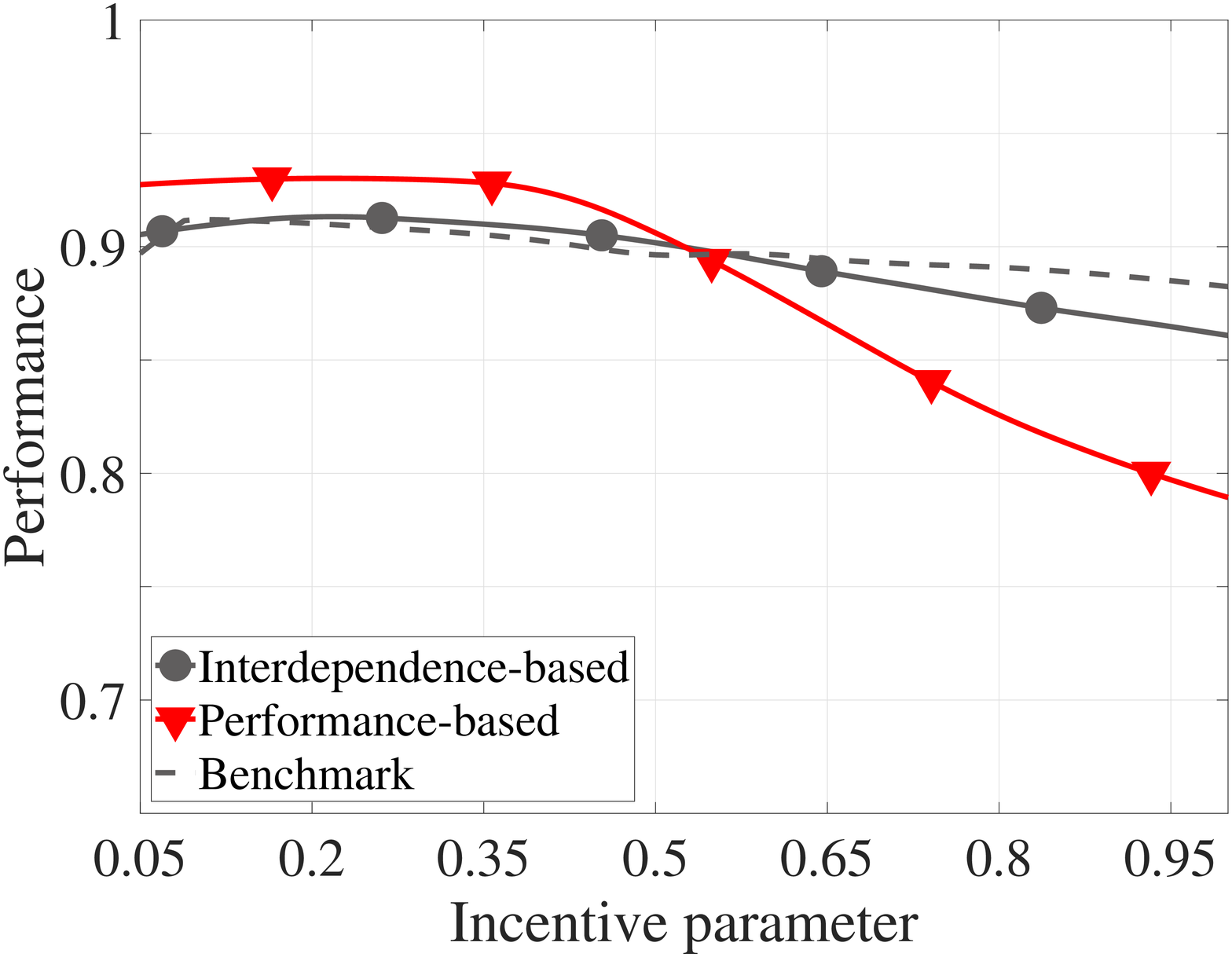}
         \label{fig:alpha-randomk4}
     \end{subfigure}
       \vspace{-4mm}
        \hspace{10mm}
    \begin{subfigure}[b]{0.33\textwidth}
         \centering
                \caption{Random pattern ($K=6$)}
         \includegraphics[width=\textwidth]{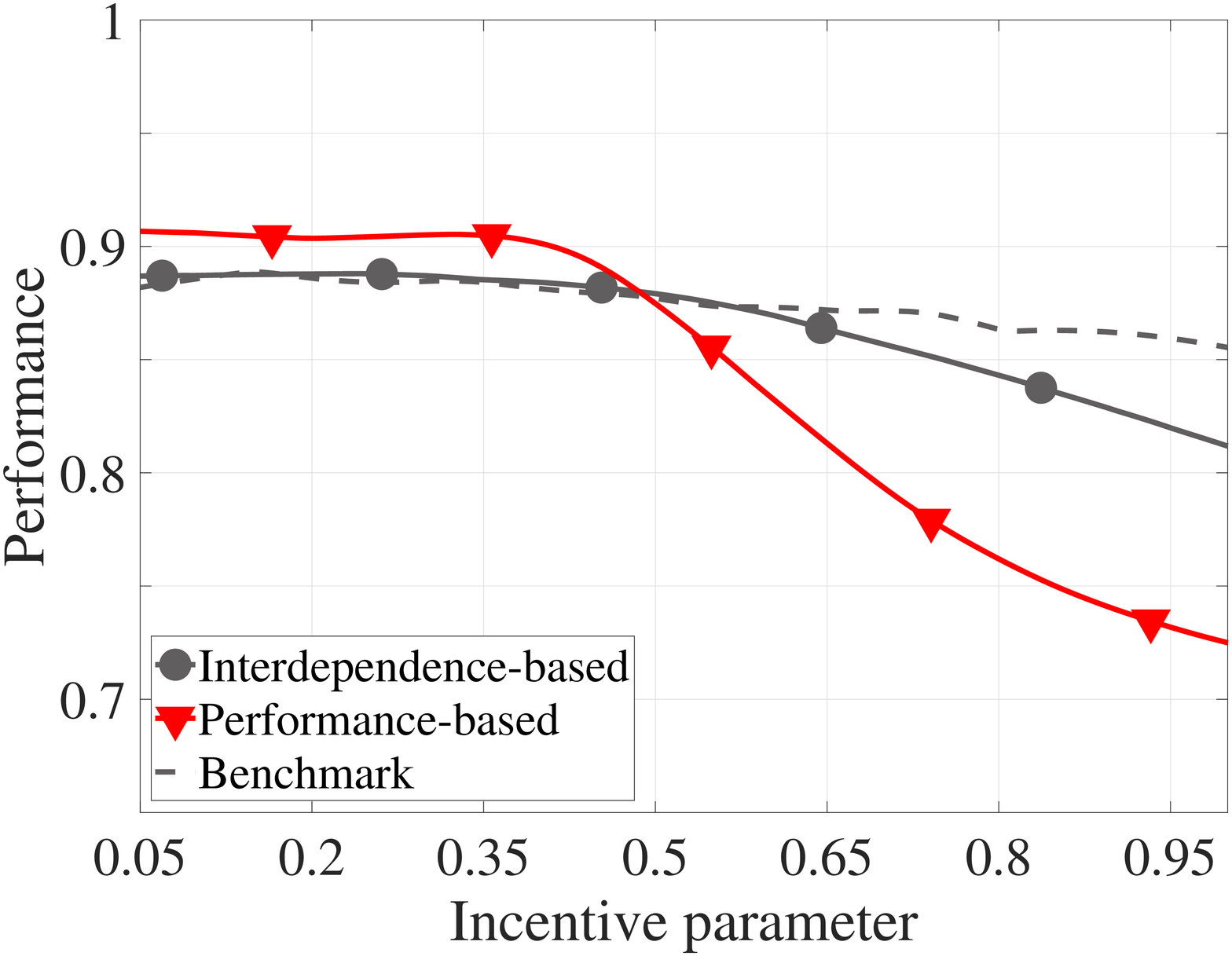}
         \label{fig:alpha-randomk6}
     \end{subfigure}
       \vspace{-4mm}

    \caption{Partial dependencies of performances on hte incentive parameter}
    \label{fig:dependency-alpha}
\end{figure}

This section examines the effects of bottom-up task allocation on performance and compares it with the more traditional top-down mode of task allocation. As introduced in Sec. \ref{sec:data-analysis}, we trained regression neural networks using the simulated data and estimated functions representing the partial dependence of the achieved performance on selected parameters. The corresponding functional dependency of the organizational performance (y-axis) on the incentive parameter (x-axis) is plotted in Fig. \ref{fig:dependency-alpha} for all analyzed interdependence patters. The red lines (triangles) and black lines (circles) represent the scenarios in which agents behave according to the performance-based and interdependence-based strategies, respectively. The benchmark scenarios in which the organizational structure is (exogenously) fixed top-down are indicated by dashed lines. It is worth noticing that relatively low values for the incentive parameter (on the left on the x-axes) indicate altruistic incentive schemes that put more emphasis on residual performance, while high values of the incentive parameter (on the right of the x-axes) result in very individualistic incentive schemes that place a stronger emphasis on the performance coming from the agents' areas of responsibility (see Eq. \ref{eq:utility}).  

The effectiveness of top-down, interdependence-based, and performance-based task allocation strategies depends on the incentive mechanisms in place within an organization. When incentive mechanisms are relatively altruistic (on the left of the x-axis), the performance-based strategy consistently outperforms the other strategies, regardless of the interdependence patterns. In terms of organizational performance, the benchmark case and interdependence based strategies perform equally well. These findings are robust across all interdependence patterns, although they are most pronounced for more complex and non-decomposable tasks (as shown in Figs. \ref{fig:alpha-5mal3reciprocal} to \ref{fig:alpha-randomk6}) and least pronounced for decomposable and almost decomposable tasks (as shown in Figs. \ref{fig:alpha-5mal3} and \ref{fig:alpha-3mal5}).

As we move from altruistic to individualistic incentive mechanisms (to the right on the x-axis), the performance of scenarios with bottom-up task allocation decreases. Interestingly, the results achieved with the performance-based strategy appear to be most sensitive to variations in the incentive parameter. As a result, performance-based task allocation---which is the superior strategy when altruistic incentives are effective---becomes the inferior strategy when incentives are more individualistic. For all interdependence structures, the shift in which  task allocation strategy is to be preferred, can be observed at intermediate values of the incentive parameter. The performance of the interdependence-based task allocation strategy also decreases, but to a significantly lesser extent than the performance-based strategy.

It is important to note that for cases with very individualistic incentive mechanisms, the benchmark case outperforms the other approaches in almost all cases. However, the results also indicate that higher levels of performance can be achieved if incentives are balanced or altruistic, and if the task is complex enough to have externalities between allocated tasks (as shown in Figs. \ref{fig:alpha-5mal3reciprocal} to \ref{fig:alpha-randomk6}). In contrast, for decomposable and almost decomposable tasks (Figs. \ref{fig:alpha-5mal3} and \ref{fig:alpha-3mal5}), the incentive parameter does not have a strong effect on performance. 

\subsection{Modularity and bottom-up task allocation}
\label{sec:results-modularity}

This section investigates the effects of independent parameters on the agents' behavior during task allocation. Examining these effects is crucial for understanding the dynamics of bottom-up task allocation. By studying how these parameters impact the agents' behavior, we can gain insights into how self-organization can be effectively guided to enhance organizational performance.

\begin{figure}
     \centering
     \begin{subfigure}[b]{0.3\textwidth}
         \centering
              \caption{Interdependence patterns}
         \includegraphics[width=\textwidth]{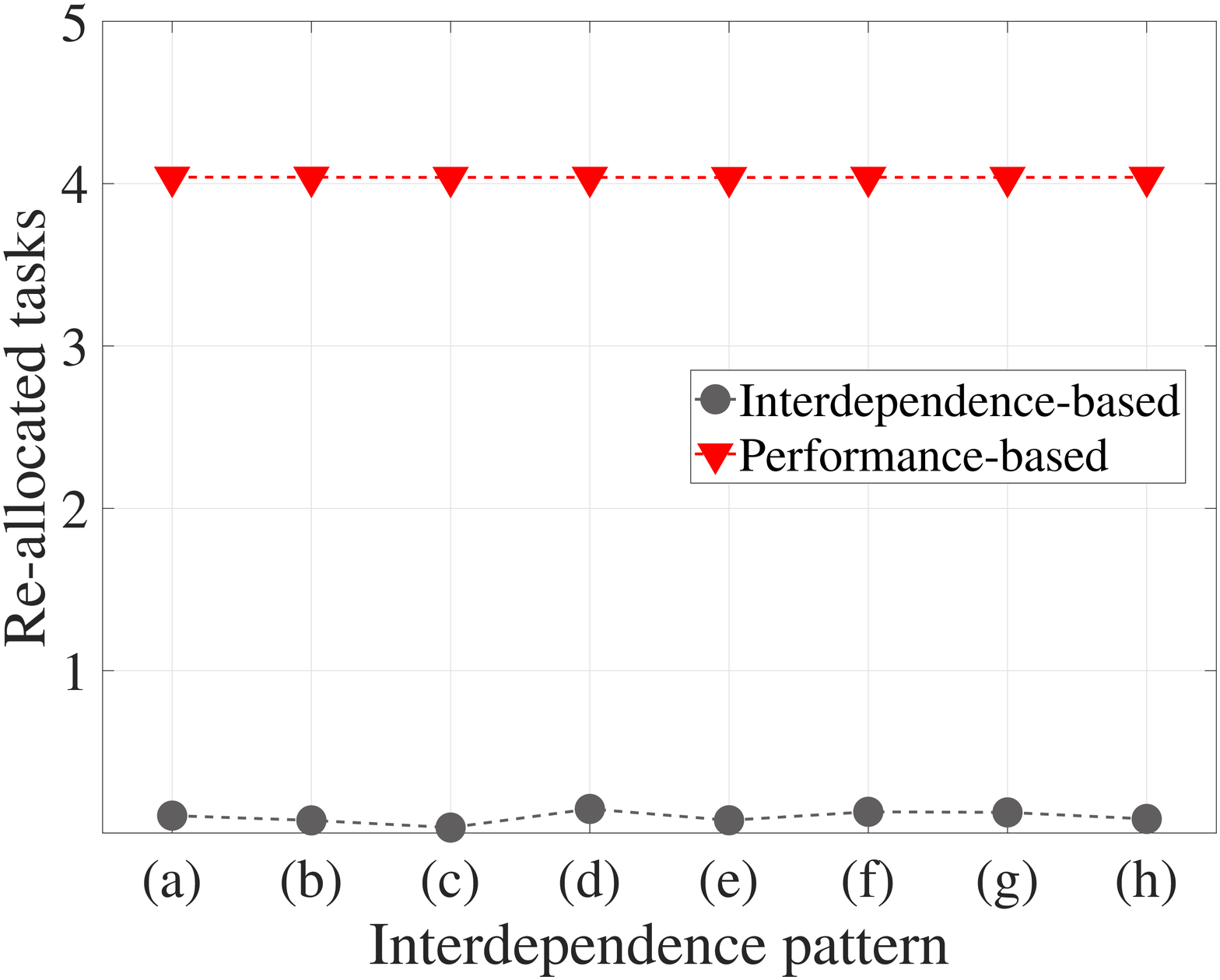}
         \label{fig:swap-complexity}
     \end{subfigure}
      \hspace{10mm}
       \vspace{-4mm}
    \begin{subfigure}[b]{0.3\textwidth}
         \centering
                  \caption{Periods with re-allocation}
         \includegraphics[width=\textwidth]{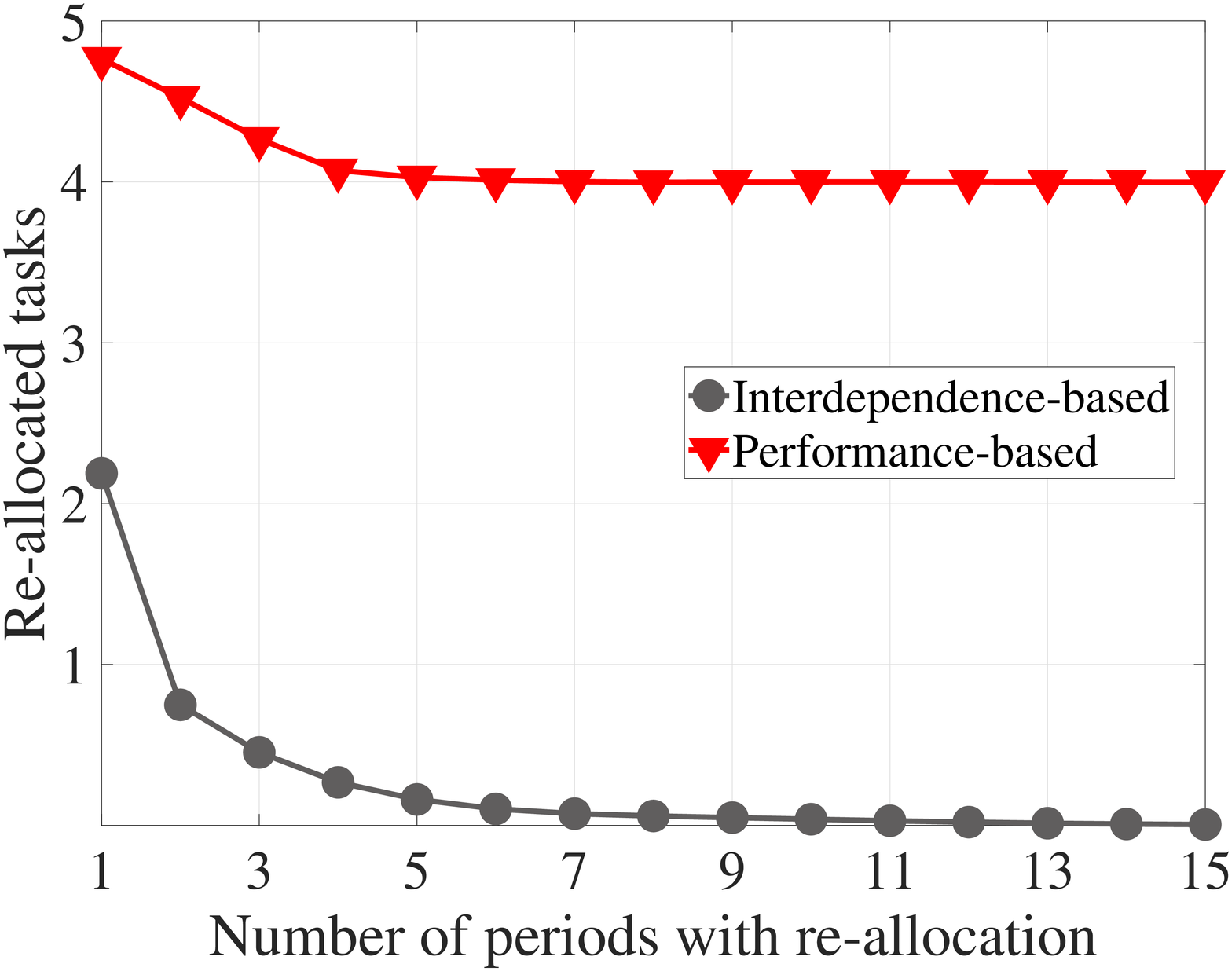}
         \label{fig:swap-rounds}
     \end{subfigure}
      \vspace{-4mm}
     
    \begin{subfigure}[b]{0.3\textwidth}
         \centering
                  \caption{Periods between re-allocation}
         \includegraphics[width=\textwidth]{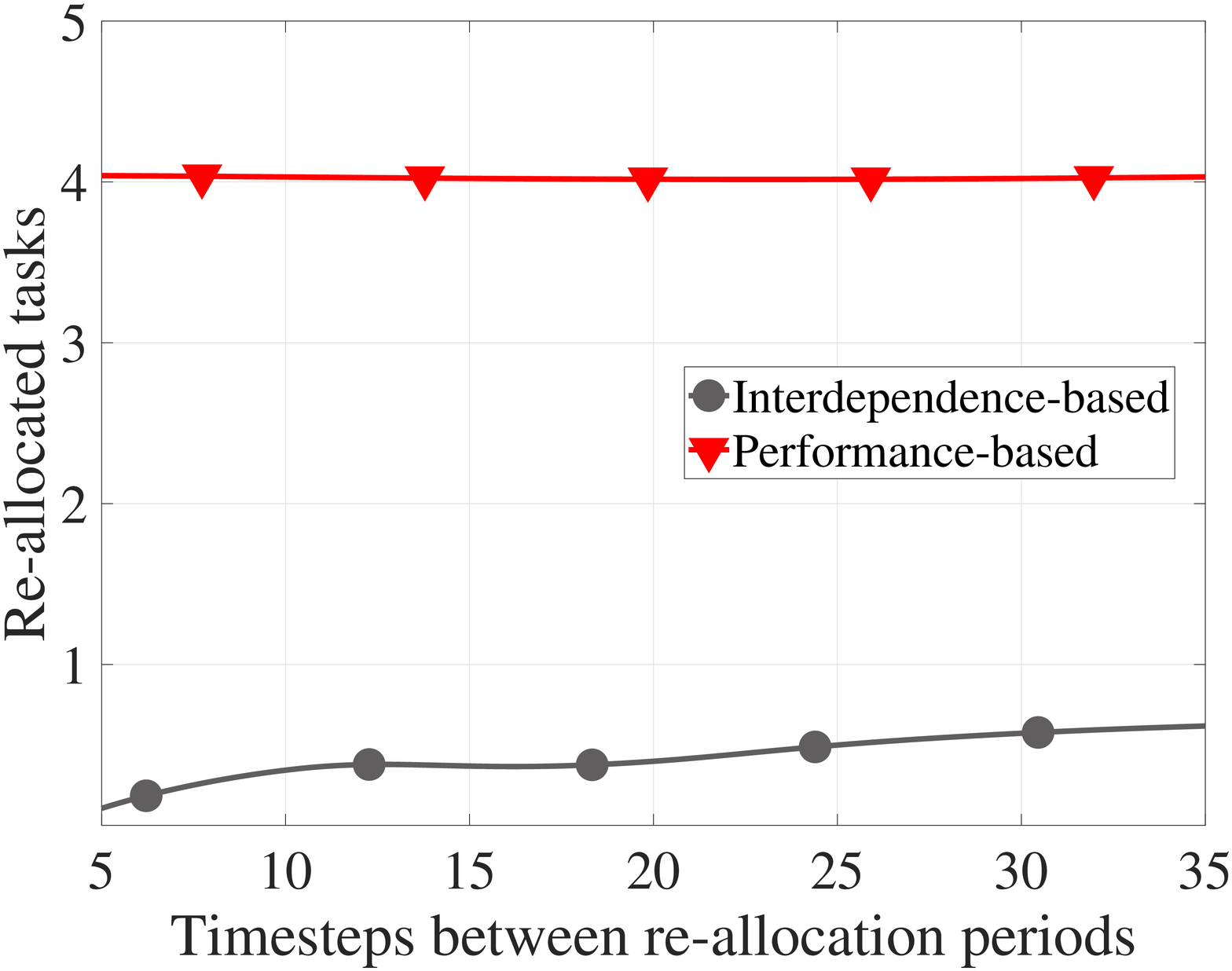}
         \label{fig:swap-tau}
     \end{subfigure}
       \hspace{10mm}
      \vspace{-4mm}
    \begin{subfigure}[b]{0.3\textwidth}
         \centering
              \caption{Incentive parameter}
         \includegraphics[width=\textwidth]{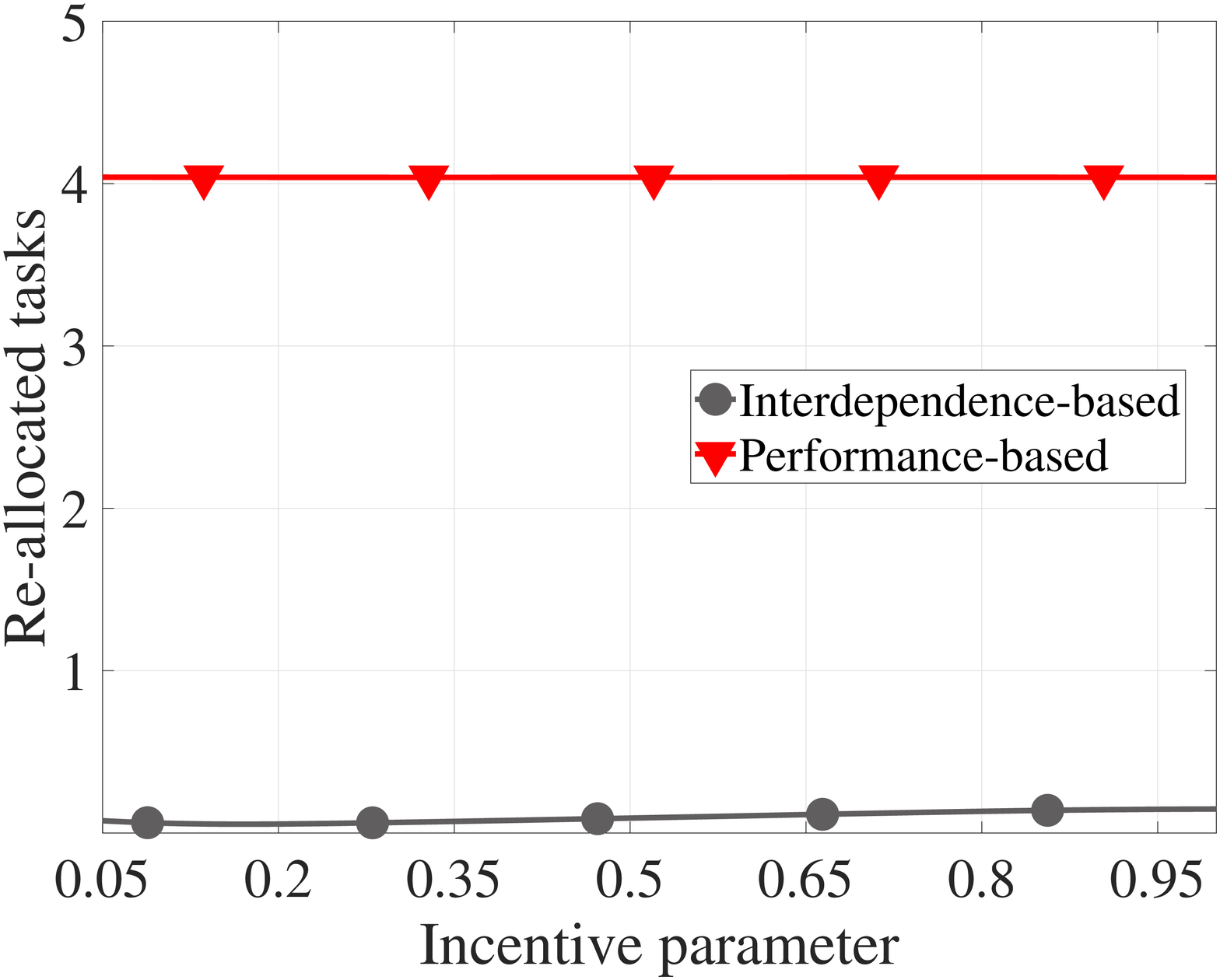}
         \label{fig:swap-alpha}
     \end{subfigure}
      \vspace{-4mm}
   
    \caption{Partial dependence of the number of re-allocated tasks on selected parameters}
    \label{fig:dependency-swaps}
\end{figure}

Figure \ref{fig:dependency-swaps} illustrates the average number of tasks exchanged when agents are allowed to reallocate tasks. As shown in the figure, the red lines (triangles) represent scenarios where agents follow a performance-based strategy, while the black lines (circles) depict cases where agents base their reallocation decisions on interdependencies. The results suggest that task allocation is more dynamic when agents follow the performance-based strategy, as evidenced by the stabilization of the number of exchanged tasks at a value of $4$ after about $4$ periods of reallocation. In contrast, the interdependence-based strategy results in a lower number of exchanged tasks, which eventually converges to zero after approximately $7$ periods of reallocation. Additionally, the results indicate that the interdependence pattern (Fig. \ref{fig:swap-complexity}), the incentive parameter (Fig. \ref{fig:swap-alpha}), and the number of periods between task allocation (Fig. \ref{fig:swap-tau}) do not or only marginally affect the average number of exchanged tasks. However, a slightly higher number of tasks are reallocated when agents have more time to learn about interdependencies (Fig. \ref{fig:swap-tau}).

\begin{figure}
     \centering
     \begin{subfigure}[b]{0.33\textwidth}
         \centering
         \caption{Small diagonal blocks ($K=2$)}
         \includegraphics[width=\textwidth]{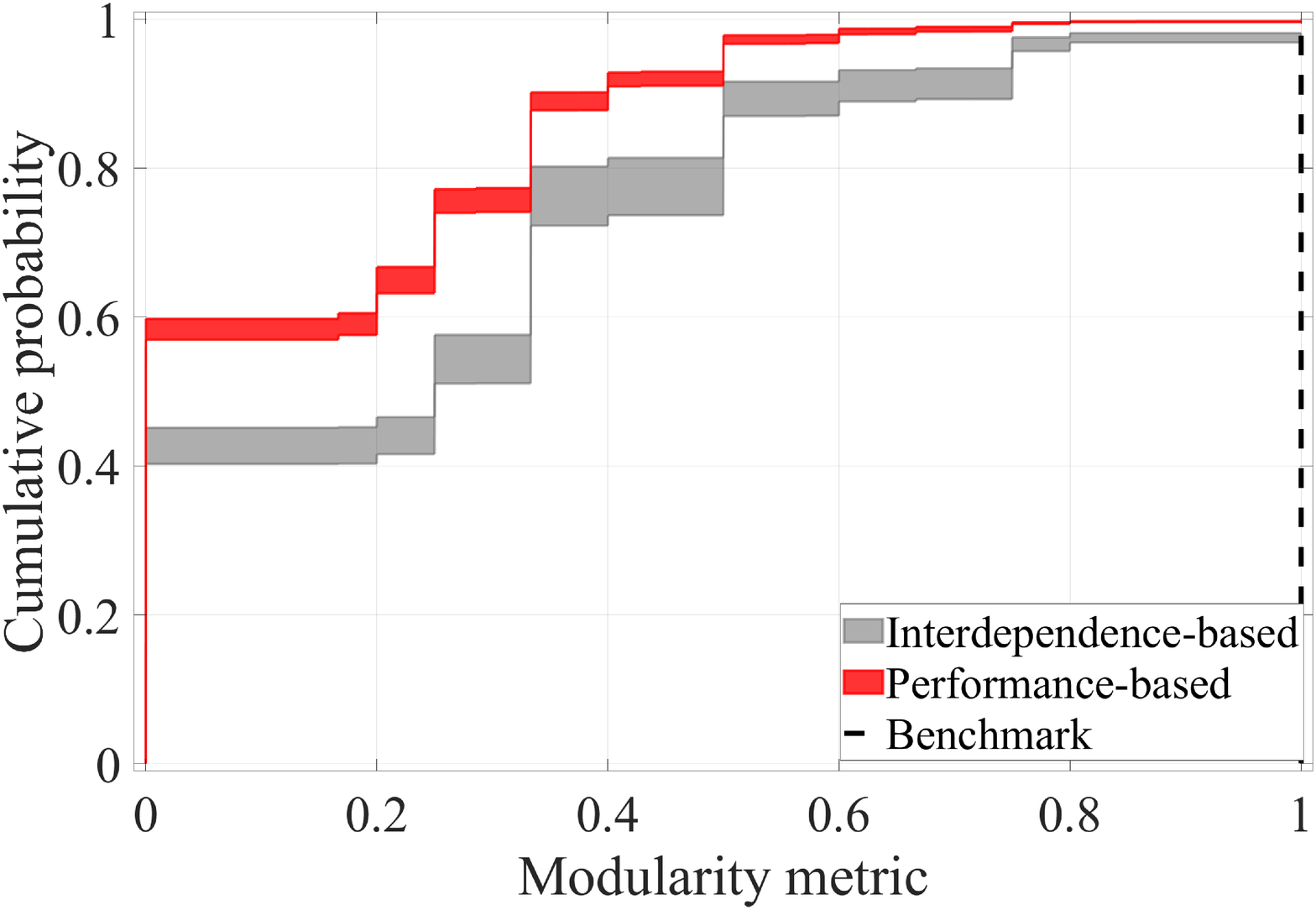}
         \label{fig:ecdf-5mal3}
     \end{subfigure}
       \hspace{10mm}
       \vspace{-4mm}
    \begin{subfigure}[b]{0.33\textwidth}
         \centering
            \caption{Big diagonal blocks ($K=4$)}
         \includegraphics[width=\textwidth]{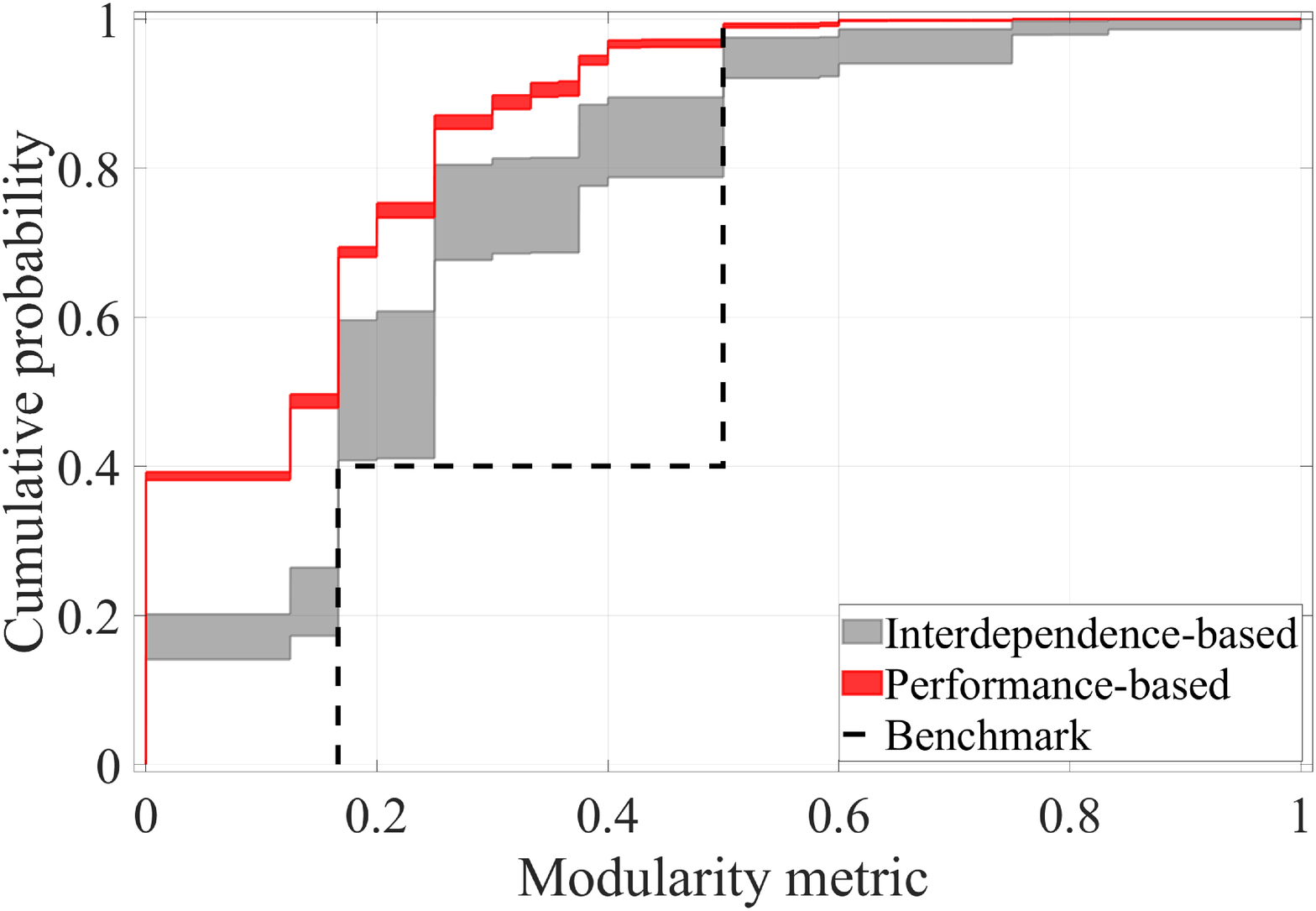}
         \label{fig:ecdf-3mal5}
     \end{subfigure}
     \vspace{-4mm}
     
    \begin{subfigure}[b]{0.33\textwidth}
         \centering
              \caption{Small blocks: Reciprocal ($K=6$)}
         \includegraphics[width=\textwidth]{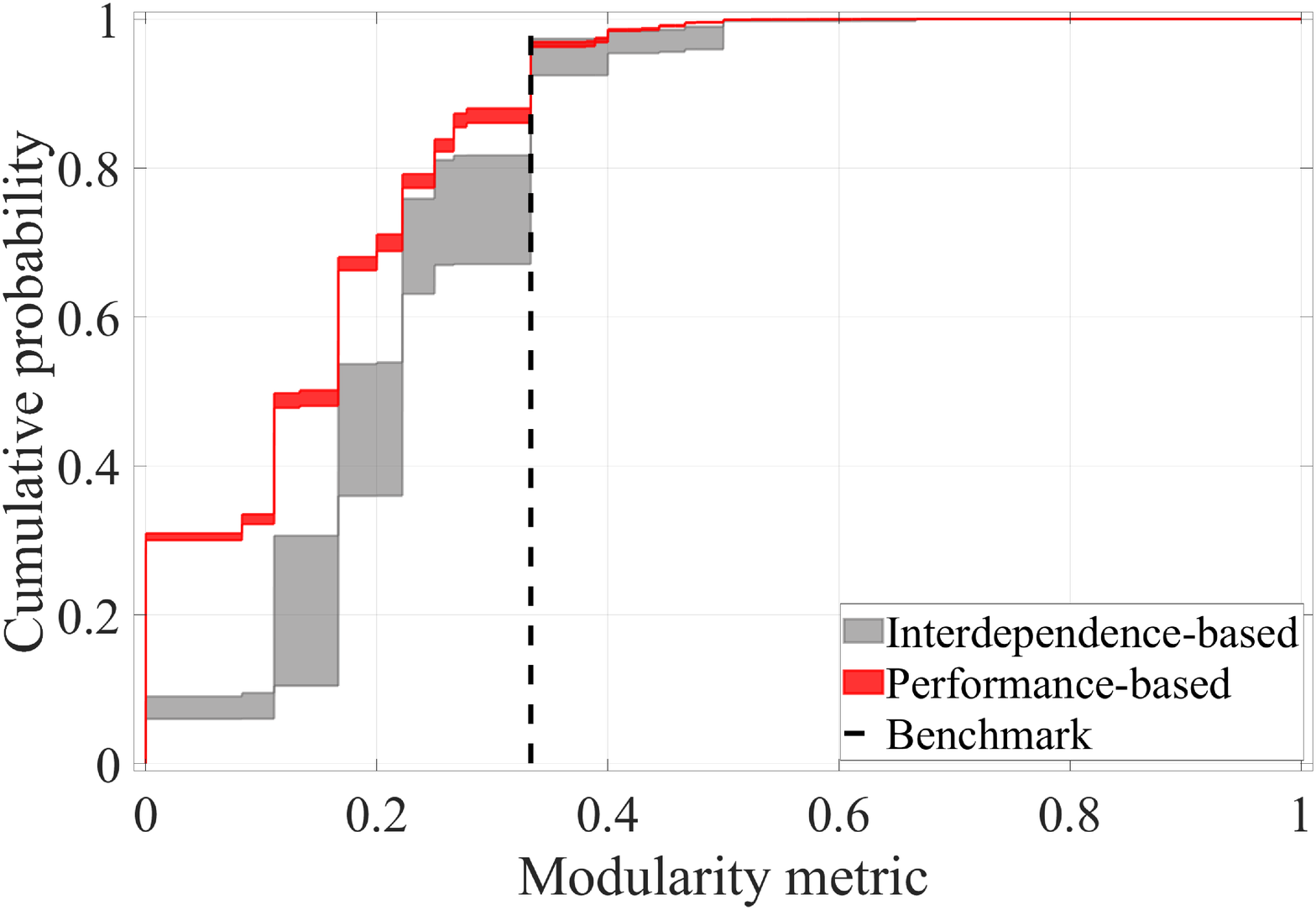}
         \label{fig:ecdf-5mal3reciprocal}
     \end{subfigure}
       \hspace{10mm}
       \vspace{-4mm}
    \begin{subfigure}[b]{0.33\textwidth}
         \centering
                  \caption{Big blocks: Reciprocal ($K=6$)}
         \includegraphics[width=\textwidth]{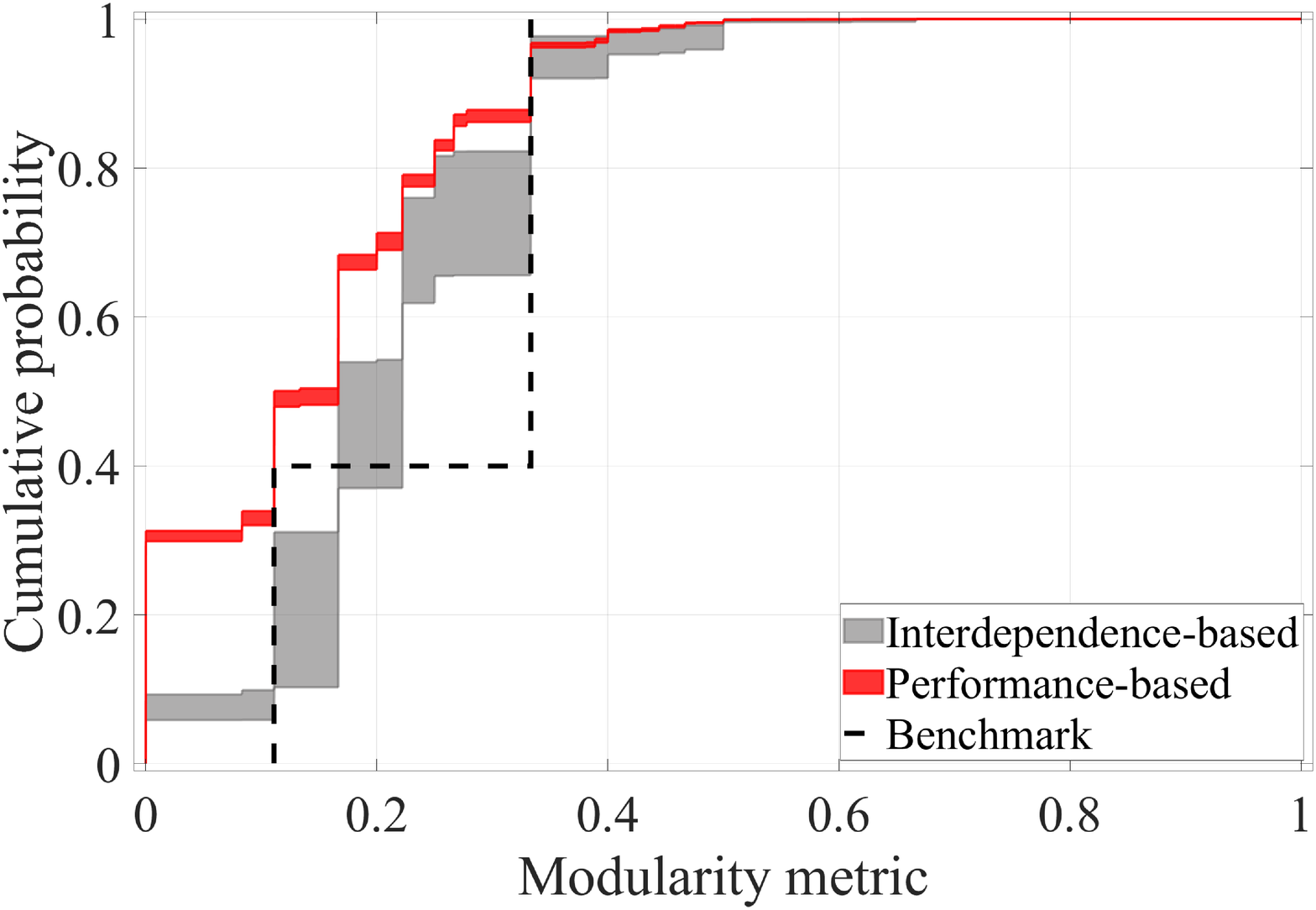}
         \label{fig:ecdf-3mal5reciprocal}
     \end{subfigure}
     \vspace{-4mm}
     
    \begin{subfigure}[b]{0.33\textwidth}
         \centering
                  \caption{Small blocks: Ring ($K=5$)}
         \includegraphics[width=\textwidth]{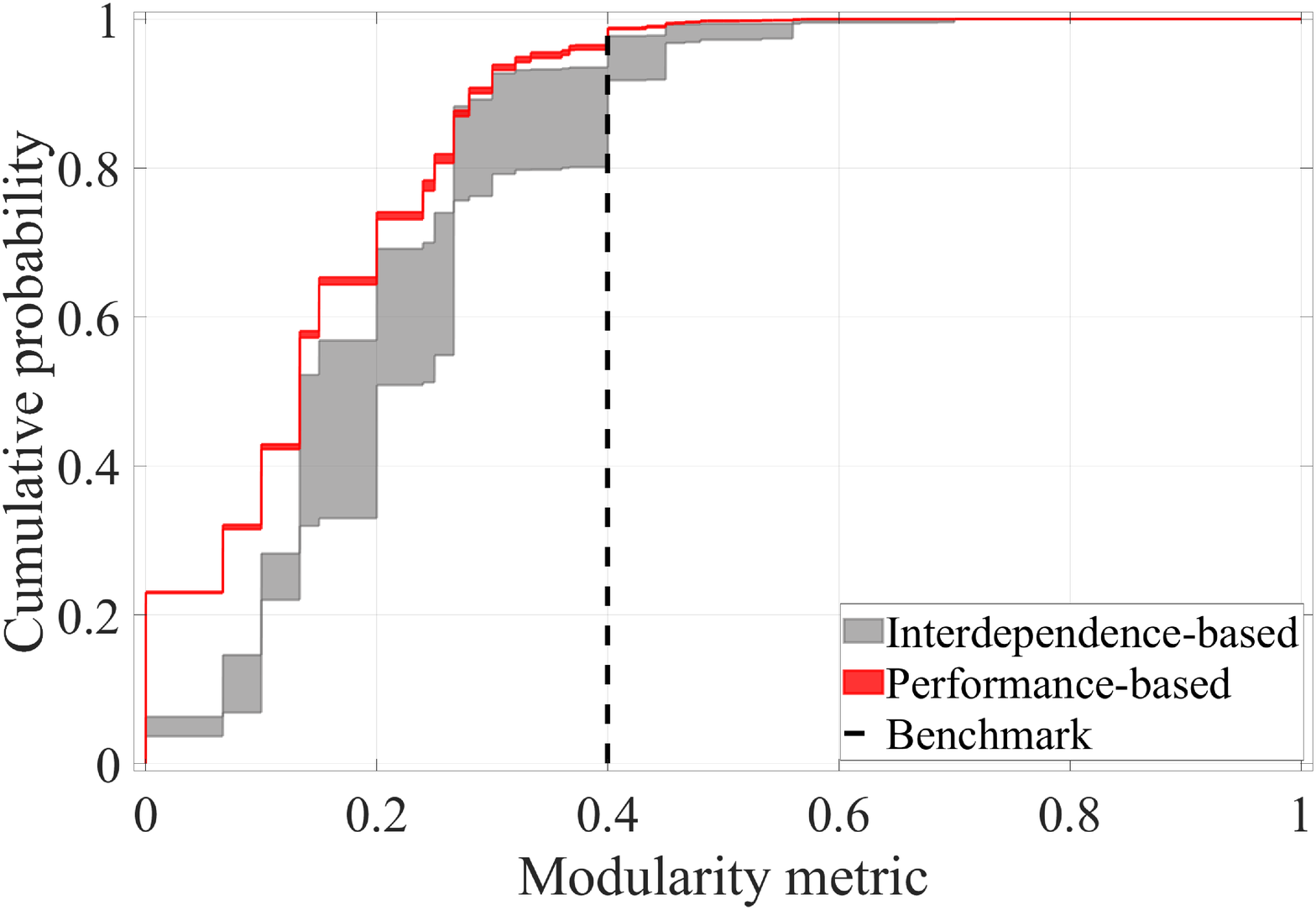}
         \label{fig:ecdf-5mal3ring}
     \end{subfigure}
       \hspace{10mm}
       \vspace{-4mm}
    \begin{subfigure}[b]{0.33\textwidth}
         \centering
           \caption{Big blocks: Ring ($K=9$)}
         \includegraphics[width=\textwidth]{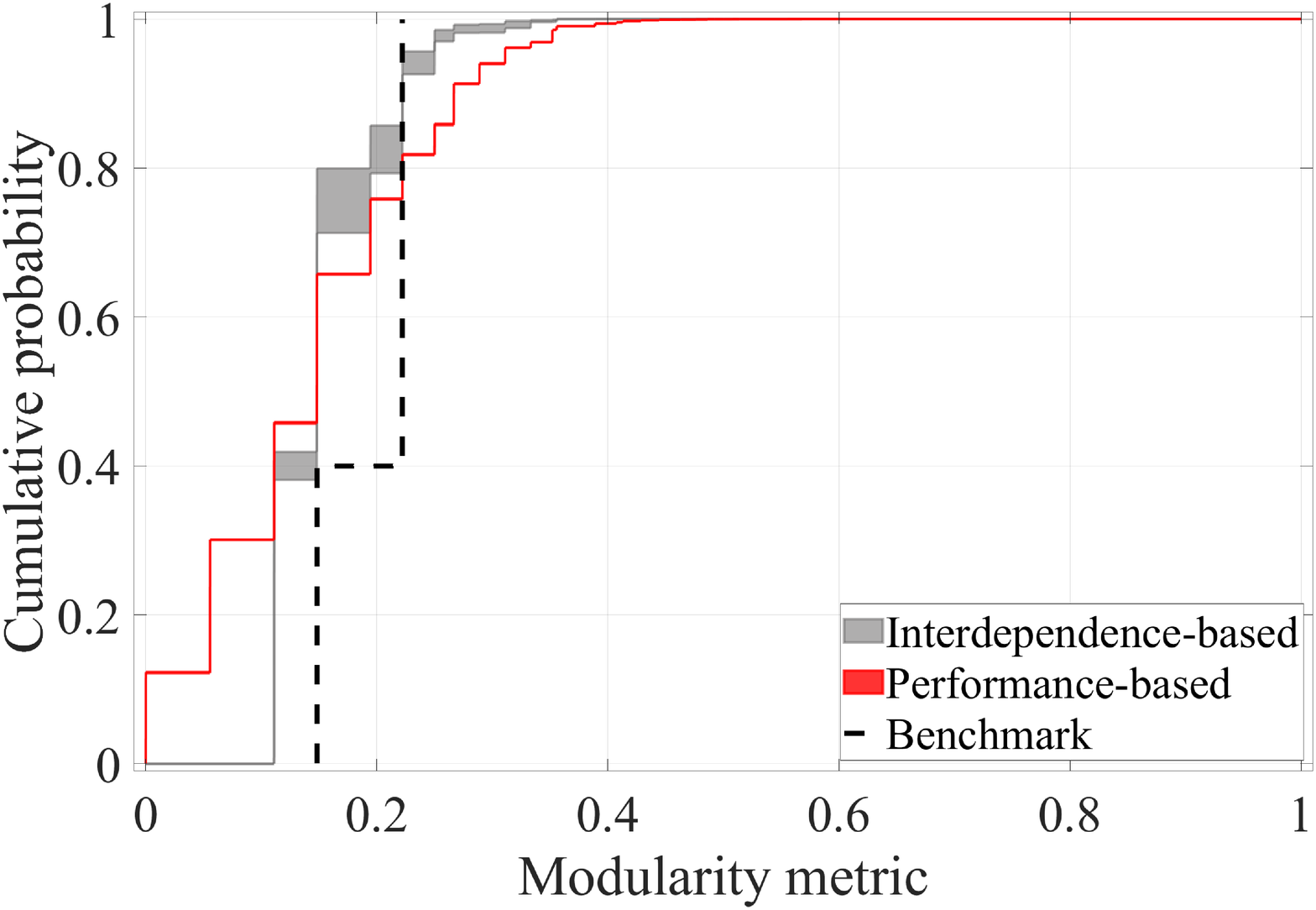}
         \label{fig:ecdf-3mal5ring}
     \end{subfigure}
     \vspace{-4mm}
     
    \begin{subfigure}[b]{0.33\textwidth}
         \centering
                  \caption{Random pattern ($K=4$)}
         \includegraphics[width=\textwidth]{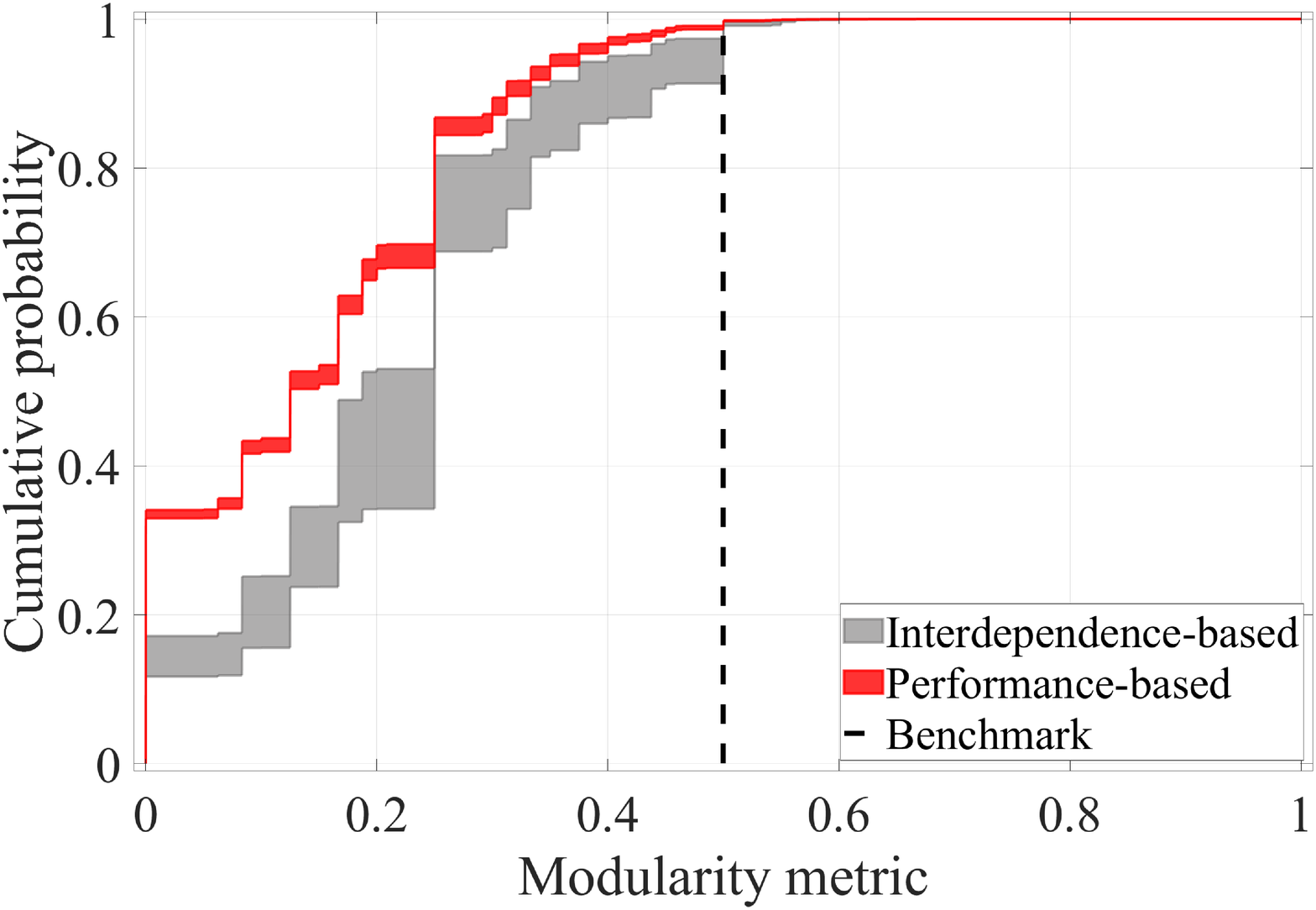}
         \label{fig:ecdf-randomk4}
     \end{subfigure}
       \hspace{10mm}
       \vspace{-4mm}
    \begin{subfigure}[b]{0.33\textwidth}
         \centering
                  \caption{Random pattern ($K=6$)}
         \includegraphics[width=\textwidth]{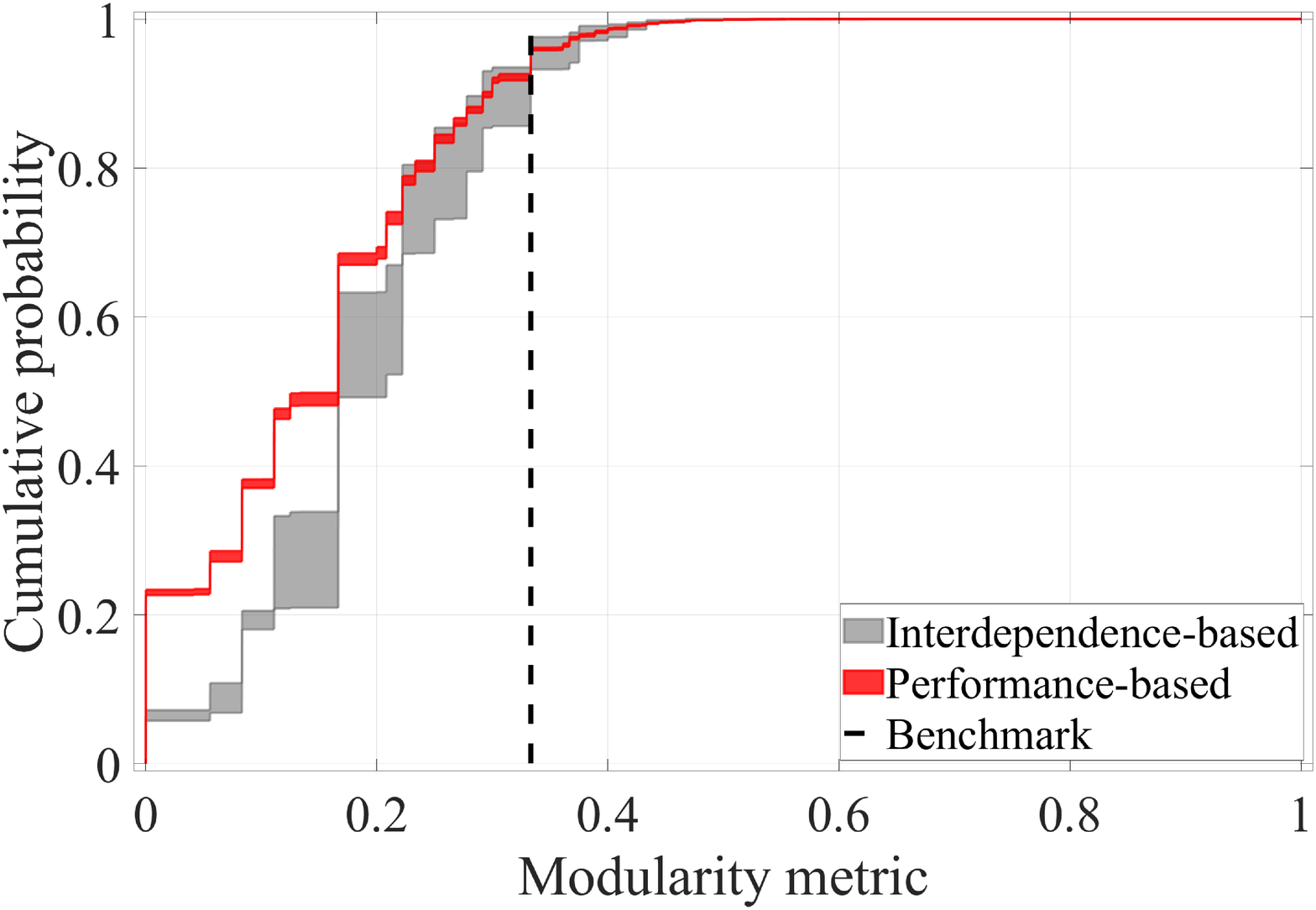}
         \label{fig:ecdf-randomk6}
     \end{subfigure}
     \vspace{-4mm}

    \caption{Cumulative distributions of modularity metrics}
    \label{fig:modularity-metric}
\end{figure}

Figure \ref{fig:modularity-metric} presents the probability distributions for the modularity metric, which measures the alignment between the organizational structure and the interdependence pattern (as defined in Eq. \ref{eq:ratio}). As discussed in Sec. \ref{sec:background}, a better alignment is expected to result in higher organizational performance. Each subplot in the figure represents the distributions for the interdependence patterns introduced in Fig. \ref{fig:matrices}. The red lines indicate the performance-based task allocation strategy and the grey lines represent the interdependence-based strategy. The colored areas show the probability ranges for different incentive parameters. In all cases, altruistic incentives result in the highest degree of modularity (at the upper edge of the colored areas), while individualistic incentives lead to the lowest degree of modularity (at the lower edge of the colored areas). For example, on the left side of Fig. \ref{fig:ecdf-5mal3}, we can see that with a probability of around $45\%$, the modularity of the emergent structure is equal to or lower than $0.2$ if agents follow an interdependence-based allocation mechanism. For scenarios with a performance-based task allocation strategy, the corresponding probability is approximately $60\%$. Additionally, the grey area is wider than the red area, indicating that the modularity in the performance-based strategy is less sensitive to the incentive parameter than in the interdependence-based strategy, particularly for high values of the modularity metric.

The dashed lines in Fig. \ref{fig:modularity-metric} represent the modularity metric of the benchmark solution. It is worth noting that for more complex decision problems, the modularity of the benchmark solution decreases due to the large number of interdependencies between decisions. For instance, Fig. \ref{fig:ecdf-5mal3} shows the results for the small diagonal block pattern in Fig. \ref{fig:matrix-5mal3}. In this pattern, the benchmark solution allows for internalizing all interdependencies within the agents' areas of responsibility, resulting in a maximum modularity of $1$. On the other hand, Fig. \ref{fig:ecdf-3mal5ring} illustrates the results for the small blocks and a ring-like interdependence structure in Fig. \ref{fig:matrix-3mal5ring}. Due to the relatively high task complexity, the benchmark solution only internalizes a fraction of interdependencies within the agents' areas of responsibility, resulting in a modularity of $0.4$. Please also note that for the structures with large diagonal blocks, not all sub-problems in the benchmark solution have the same modularity, which leads to the step-like pattern of the benchmark modularity in Figs. \ref{fig:ecdf-3mal5}, \ref{fig:ecdf-3mal5reciprocal}, and \ref{fig:ecdf-3mal5ring}.

Overall, we observe that when agents follow a performance-based task allocation strategy, there is a higher likelihood of \textit{low} \textit{modularity} in the emergent task allocation pattern. On the other hand, if agents learn about interdependencies over time and optimize their task allocation based on this information, the resulting task allocation pattern is more likely to be \textit{more modular}. However, it is surprising that in most cases, the modularity of the emergent task allocation is lower than that of the benchmark solution. For example, in Fig. \ref{fig:ecdf-5mal3reciprocal}, we see that in around $20\%$ of cases with an interdependence-based approach and in around $10\%$ of cases with a performance-based task allocation, the modularity of the benchmark solution is achieved or exceeded. For most other scenarios, this pattern is even more pronounced. For instance, in the results for small diagonal blocks and ring-like interdependence patterns, the benchmark modularity is achieved in less than $10\%$ of cases.

\subsection{Discussion}

The results presented in the previous sections are summarized and brought in the context of the research questions that were defined in the introduction in Tab. \ref{tab:summary_results}. The results show that basing task reallocation on performance considerations tends to lead to less modular structures, while allowing agents to learn about interdependencies over time and base their task allocation on this learned information results in slightly more modular structures. Previous research on organizational design has yielded mixed results regarding the mirroring hypothesis, with some studies supporting it and others rejecting it (see Sec. \ref{sec:background}). The results presented in this paper provide further insight into this topic and may help to explain the ambiguous findings in previous research.

The findings of this study suggest that when designing organizations for bottom-up task reallocation, mirroring should be considered in a broader context, taking into account factors such as internal fit with other design elements, such as incentive mechanisms and task uncertainty. Previous research has demonstrated that an internal fit between routines for autonomous task reallocation and other design elements, as well as task uncertainty, can impact organizational performance \citep{donaldson2014, schlevogt2002, burns1961, lawrence1967}. In the model presented in this study, task uncertainty arises from the interdependencies between individual decisions, leading to uncertain decision-making outcomes for each agent. More complex tasks result in greater uncertainty. It should be noted that organizational design involves many complex choices beyond the elements considered in this paper.\footnote{For comprehensive discussions on the complexity of organizational design choices, the reader is referred to \cite{puranam2012} and \cite{caspin2013} and the literature referenced therein.} Optimizing for a fit across all design elements is infeasible \citep{simon1967}. Therefore, in line with the idea of considering autonomous task allocation as a self-organization process and guiding it towards a desired outcome \citep{prokopenko2009}, this paper focuses on the interplay between bottom-up task allocation routines, the incentive mechanisms that control behavior, and organizational performance, particularly in relation to the resulting modularity of the structure.

The following two sections explore the emerging dynamics. In particular, Sec. \ref{sec:discussion-altruistic-incentives} examines situations with uncertainty due to complex tasks and explains why there is a good fit between altruistic incentives and performance-based task reallocation mechanisms based on the results and previous research. Section \ref{sec:discussion-individualistic-incentives} focuses on cases with individualistic incentive mechanisms and highlights the importance of considering mirroring in order to mitigate negative impacts of misfit in organizational design.

\begin{table}[htp]
\begin{scriptsize}
\caption{Summary of results}
\label{tab:summary_results}
\begin{tabular}{p{0.32\textwidth}p{0.62\textwidth}}
Research Question & Results\\ 
\noalign{\smallskip} 
\hline 
\noalign{\smallskip} 
\begin{enumerate}[leftmargin=0.5cm,nosep]
\item[(1)] How do bottom-up task allocation mechanisms and incentive mechanisms interact and influence organizational performance? 
\end{enumerate}
&
\begin{itemize}[leftmargin=0.4cm,nosep,noitemsep,nolistsep]
\item The highest performances are observed when the organization uses altruistic incentives and a performance-based approach to task re-allocation.
\item If individualistic incentives are present, the best performances are achieved through interdependence-based or top-down task allocation.
\item The sensitivity of performance to the incentive parameter is greatest when task allocation is based on performance.
\end{itemize}
\vfill
\\
\noalign{\smallskip} \hline \noalign{\smallskip}
\begin{enumerate}[leftmargin=0.5cm,nosep]
\item[(2)] To what extent is the structure that emerges from bottom-up task allocation modular and how does this affect organizational performance? 
\end{enumerate}
&
\begin{itemize}[leftmargin=0.4cm,nosep,noitemsep,nolistsep]
\item The modularity achieved by the emergent structure falls below the benchmark solution.
\item Altruistic incentive mechanisms tend to produce slightly more modular structures compared to individualistic incentives.
\item Interdependence-based strategies result in higher modularity compared to performance-based strategies.
\item Modularity is particularly important when individualistic incentive mechanisms are in place.
\end{itemize}
\\
\noalign{\smallskip} \hline \noalign{\smallskip}
\begin{enumerate}[leftmargin=0.5cm,nosep]
\item[(3)] How does task complexity impact the results observed in relation to the interactions between bottom-up task allocation and incentive mechanisms and modularity?
\end{enumerate}
&
\begin{itemize}[leftmargin=0.4cm,nosep,noitemsep,nolistsep]
\item The impact of incentive mechanisms on organizational performance is more pronounced for tasks with high complexity.
\item The sensitivity of performance to the incentive parameter is stronger for tasks with high complexity
\item The results are robust across interdependence structures.
\end{itemize}
\\
\noalign{\smallskip} \hline
\end{tabular}
\end{scriptsize}
\end{table}

\subsubsection{Altruistic incentive mechanisms and bottom-up task allocation}
\label{sec:discussion-altruistic-incentives}

The findings in Sec. \ref{sec:results-performance} reveal that combining altruistic incentives with performance-based task allocation strategies is particularly effective in situations where tasks are complex and interdependent. This may seem counter-intuitive, as one might assume that self-oriented and myopic behavior from agents would not result in improved performance for the organization. However, our results suggest that this may not always be the case.

The micro-level dynamics of combined selfish and altruistic behavior may help to shed light on these findings. Previous research has suggested that individual behavior is often influenced by both selfish and altruistic motivations. For instance, on online platforms that feature user-generated content, people may engage in behaviors such as contributing information in an effort to gain reputation and recognition, as well as to help others \citep{hennig2004,daugherty2008,qiao2017}. It is often assumed that altruistic motives are primarily intrinsic, but this model suggests that such behavior may actually be motivated by a desire to maximize utility and the expectation of reciprocation from others \citep{gneezy2011,deci1971}. 
The altruistic incentive mechanism places a great deal of emphasis on the performance contributions of other individuals. As a result, an individual may align their actions with those of others in an attempt to increase \textit{their own} utility and may expect similar behavior in return, a concept known as reciprocity \citep{fehr2000,fehr2006,falk2006}. Thus, while the mutual support between individuals may appear to be prosocial in nature, it may actually be motivated by self-interest rather than genuine altruism.

The results suggest that the performance-based task reallocation mechanism promotes the efficient allocation of tasks by ensuring that each agent is responsible for the tasks that they can perform best. This finding may be due to the mutually reinforcing mechanisms of an incentive mechanism that encourages agents to support each other by aligning their decisions and a task reallocation mechanism that assigns each agent the tasks that they can perform best, given the mutual support between agents. Additionally, our results indicate that these mechanisms are particularly effective when there are interdependencies between the agents' areas of responsibility.

The dynamic nature of bottom-up task allocation may also contribute to the effectiveness of altruistic incentive mechanisms and performance-based task reallocation in complex tasks. As shown in Fig. \ref{fig:dependency-swaps}, the performance-based strategy results in frequent swaps of tasks between agents, which creates a mechanism similar to job rotation. In this process, agents take on different tasks within the organization and, thanks to the performance-based reallocation mechanism, eventually work on the tasks they are best suited for. By combining altruistic incentive mechanisms with performance-based task reallocation, we not only address the job assignment problem identified in \cite{ortega2001}---which suggests that organizations can learn about employees through job rotation \citep[see also][]{jovanovic1979,meyer1994}---but we also allow agents to \textit{learn for themselves} which tasks they can perform best. This knowledge is then used to determine the division of labor, and the altruistic incentives ensure coordinated action across organizational units, ultimately improving organizational performance.

\subsubsection{Individualistic incentive mechanisms and bottom-up task allocation}
\label{sec:discussion-individualistic-incentives}

As soon as tasks are sufficiently complex so that they are no longer decomposable, organizations are worse off if they employ individualistic incentives \citep{fischer2008}. The model introduced in this paper replicates this finding and additionally generates insights into the dynamics of bottom-up task allocation in the case of individualistic incentives. Contrary to the findings discussed in Sec. \ref{sec:discussion-altruistic-incentives}, individualistic incentives work best with either top-down task allocation or interdependence-based bottom-up task approaches to task allocation, while the organization's performance is substantially lower if task allocation follows a performance-based approach. 

Internalizing interdependencies within the areas of responsibility of organizational units can reduce the need for behavioral control to prevent selfish behavior among agents \citep{foss2015}. This approach aligns with the recommendations given by transaction cost economics, which usually assumes selfish behavior on the part of agents, similar to the behavior induced by individualistic incentive mechanisms \citep{williamson1985,boyacigiller199,erez1987}. Top-down approaches to task allocation follow this logic by internalizing a large number of interdependencies, which helps to mitigate the negative effects of individualistic incentives and leads to better organizational performance compared to bottom-up task allocation. This is supported by the results in Fig. \ref{fig:modularity-metric}, which show that bottom-up approaches do not achieve the desired modularity, while interdependence-based approaches are more modular than performance-based approaches. The achieved organizational performances follow this pattern. 
Therefore, we can conclude that either top-down or interdependence-based bottom-up approaches to task allocation can be effective in organizational design if individualistic incentives are in place. However, as discussed in Sec. \ref{sec:discussion-altruistic-incentives}, a combination of altruistic incentives mechanisms and performance-based incentives may lead to even higher performances.


\section{Summary and conclusion}
\label{sec:conclusion}

This paper presents an agent-based model of a stylized organization with bottom-up task allocation to examine the interplay between task allocation and incentive mechanisms, as well as the impact of task complexity. The results suggest that bottom-up approaches can be effective in certain circumstances, particularly when incentives are relatively altruistic and tasks are complex with externalities between allocated tasks. In these cases, performance-based task allocation outperforms both the traditional top-down approach and interdependence-based task allocation. However, as incentives become more individualistic, the performance of bottom-up approaches decreases, and the traditional top-down approach becomes the superior strategy. These findings are consistent with previous research on incentives in organizations \citep[e.g.,][]{fischer2008}. The results also indicate that the modularity of top-down designed organizations is generally not achieved through bottom-up task allocation. However, this paper also highlights that modularity only is relevant individualistic incentives are present in the organization. 

This research has several limitations that should be considered when interpreting the results. First, the model assumes that coordination occurs only through the incentive mechanism and does not account for communication between agents. While this may be realistic in some cases, future research could improve the validity of the results by incorporating communication between agents. Second, the model focuses on a single organization and does not consider external influences. Future research could examine the co-evolution of multiple interdependent organizations \citep{volberda2003} that might be organized using less hierarchical concepts, such as self-organized or self-managed forms. Finally, the analysis in this paper is based on the symmetrical $NK$-framework, which assumes that interdependencies are evenly distributed across the landscape. However, in reality, interdependencies may be concentrated, resulting in plateaued landscapes \citep{hebbron2008} that were not considered in this study.

\bibliographystyle{elsarticle-harv} 
\bibliography{bib.bib}

\end{document}